\documentclass[12pt]{article}
\usepackage[utf8]{inputenc}
\usepackage[english]{babel}
\usepackage{amsfonts}
\usepackage{textcomp}
\usepackage{amsmath, amssymb, graphicx}
\renewcommand{\theequation}{\arabic{section}.\arabic{equation}}
\usepackage{geometry}
\usepackage{enumerate}
\usepackage{hyperref}
\usepackage{braket}
\usepackage{color}
\usepackage{cite}
\usepackage{float}
\usepackage{indentfirst}
\usepackage{epsfig}
\usepackage{booktabs}
\usepackage{subfigure}
\usepackage{caption}
\usepackage{ulem}
\usepackage{array}
\usepackage{multirow}

\newcommand{\be}{\begin{equation}}
\newcommand{\ee}{\end{equation}}
\newcommand{\bea}{\begin{eqnarray}}
\newcommand{\eea}{\end{eqnarray}}

\newcommand{\fnm}{\footnotemark}
\newcommand{\fnt}{\footnotetext}

\begin{document}

\begin{center}

  \large \bf
  Stability analysis of holographic RG flows in 3d supergravity
\end{center}

\vspace{15pt}

\begin{center}

 \normalsize\bf
        Anastasia A. Golubtsova\fnm[1]\fnt[1]{golubtsova@theor.jinr.ru}$^{, a,b}$
       and  Marina K. Usova \fnm[2]\fnt[2]{usovamk@mi-ras.ru}$^{, a,b}$

 \vspace{7pt}
 
  \it (a) \ \ \ \ Bogoliubov Laboratory of Theoretical Physics, JINR,\\
Joliot-Curie str. 6,  Dubna, 141980  Russia  \\

\it (b)\ Steklov Mathematical Institute, Russian Academy of Sciences\\
Gubkina str. 8, 119991 Moscow, Russia  


 \end{center}
 
  \vspace{15pt}

\begin{abstract}		

 We study holographic RG flows in a 3d supergravity model from the side of the dynamical system theory. The gravity equations of motion are reduced to an autonomous dynamical system. 
 Then we find equilibrium points of the system and analyze them for stability.  We also restore asymptotic solutions near the critical points. We find two types of solutions: with asymptotically AdS metrics
and  hyperscaling violating metrics. We write down possible RG flows between an unstable (saddle) UV fixed point  and a stable (stable node) IR fixed point. 
We also analyze bifurcations in the model.

\end{abstract}	

\tableofcontents

\section{Introduction}

 The renormalization group approach \cite{WK} provides a  systematic picture for understanding the dynamics of systems with many degrees of freedom.  
 The renormalization group flow describes changes of a physical system as a scale of a theory varies. Encoding the dependence of coupling constants on the energy scale can be represented by beta functions of a theory \cite{WK}. The zeroes of the $\beta$ functions are identified by fixed points of RG flows. 
The definition of fixed points implies that the couplings should be independent of the energy scale. For instance, in the UV limit the QCD coupling asymptotes to the Gaussian scale invariant fixed point \cite{GW}. At the same time a theory may flow to an IR fixed point, so such theory is scale-invariant at the large distances.

  Studies of fixed points play an important role, providing information about behaviour of theories at phase transitions. It is natural to explore RG flows and fixed points using holography that allows to capture inaccessible phenomena at strong coupling in the dual gauge theory \cite{Akhmedov,BVV}.  Holographic duality gives a description of RG flows in terms of gravitational solutions of a special type for a gravity model coupled to a dilaton \cite{BVV,deB,BST, BFS, Bianchi:2001de, Papadimitriou:2004ap, Papadimitriou:2004rz, FGPW, Papadimitriou:2007sj}. Holographic RG flows can be represented by Poincaré invariant domain walls that flow between two boundaries. In turn the boundaries are associated with UV and IR fixed points of a dual field theory.  Poincaré invariance of the metric allows to pass from Einstein second order differential equations to first-order ones.
  
  It worth to be noted that a consistent formulation of RG flow equations in terms of gravity ones was precisely formulated only for solutions invariant under the Poincaré group \cite{Heemskerk:2010hk}. In this case it can be shown that the equations of motion can be rewritten in terms of the superpotential, which is defined by a scalar potential (up to the choice of some integration constants) and is in one-to-one correspondence with the $\beta$-function of the dual theory.
  
 The domain wall solutions can be associated to deformations of CFT either by relevant operators or by turning  vacuum expectation values of these operators. The asymptotic of the metric should satisfy the appropriate asymptotic behaviour  and scalar bulk fields should obey certain boundary conditions. In particular, the  Dirichlet boundary conditions
indicate that the derivative with respect to the radial (holographic) variable is asymptotically identified with the dilatation operator of the dual field theory \cite{Papadimitriou:2004ap, Papadimitriou:2004rz}.
Then one can represent the holographic $\beta$-function as follows
 \be
 \label{betaf}
 \beta = \frac{d\lambda}{d \mathcal{A}},\nonumber
 \ee
 where  $\mathcal{A}=e^{A}$ and  $\lambda=e^{\phi}$ with the scale factor $A$  and the scalar field $\phi$.
 
 Note that to give a domain wall solution an interpretation of holographic RG flow, the scale factor of the solution should be a monotonic function and the scalar field potential is restricted from above $V(\phi)\leq0$, i.e. it should obey the Gubser bound \cite{Gubser:2000nd}.

 The study of RG flows using the language of dynamical system naturally occurs thanking to $\beta$-function equations. RG flow equations represent first-order differential equations. 
The equilibrium points of these differential equations are identified with fixed points of RG flows. A connection between RG flows and dynamical systems was considered in \cite{Gukov, UG}. It is important to investigate stability of fixed points since it's helpful to find out the destiny of the system. Infrared fixed points control the long distance and low momentum behaviour of theories. The knowledge of stability of IR fixed points can give an information on phase transitions. The bifurcation analysis can shed a light if any operator crosses through marginality in RG flows.

 Holographic RG flow were widely investigated using the perturbation analysis near extremal points of a generic type potential/superpotential in works \cite{Gursoy:2008za}-\cite{Ghosh:2018qtg}. It was presented holographic RG flows with non-trivial behaviour, that was recently observed from the quantum field theory side. Using a gravitational integrable model holographic RG flows were studied in \cite{AGP}. In \cite{ARRA,IAHoReGR,IYaRa,Cartwright:2021hpv} a generalization of  RG flow equations to a holographic system with a chemical potential was considered.

   Analytic domain wall solutions in 3d gauged supergravities, which correspond to holographic RG flows,  were found in \cite{Berg:2001ty,Deger:2002hv,Chatrabhuti:2010dh}. In \cite{ParkRoLee} it was discussed a holographic RG flow from an $AdS_3$ UV fixed point to an $AdS_{3}$ IR fixed point for the model suggested in \cite{Deger:2002hv}, the central charges were also calculated.

The aim of this work is to explore holographic RG flows in a 3d supergravity model with a scalar field  and its potential \cite{Deger:2002hv,Sezgin} in terms of the dynamical system theory. Introducing new variables we rewrite equations of motion of the holographic model as a set of autonomous differential equations. We find equilibrium points of the system and explore their stability. Then we reconstruct asymptotic solutions for the metric and the scalar field near these critical points. The solutions, that  satisfy EOM and have a monotonic scale factor, can be interpreted as  fixed points of a dual field theory. We find that asymptotically AdS solutions can describe  UV fixed points, which are unstable (saddles).

The outline of this paper is as follows. In Section 2 we briefly describe the holographic model, we write down equations of motion  for the domain wall ansatz  and discuss an exact solution from \cite{Deger:2002hv}. In Section 3 scaling dimensions of operators are calculated. In Section 4, the equations of motion are reduced to an autonomous dynamical system of differential equations, for which fixed points are found and phase portraits are presented. In this section we also explore stability of the fixed points.  In Section 5, asymptotic solutions for the scalar field and the metric are reconstructed. Then we check the consistency of the asymptotic  solutions with the gravitational equations of motion. Finally, we present examples of RG flows in the model.
We also discuss  bifurcations in the model using eigenvalues of critical points and scaling dimensions of operators.
We give discussion of our results in Section 6.

\section{3d $\mathcal{N}=2$ supergravity model}
	
In \cite{Sezgin} a 3d gauged supergravity model was constructed with a  multiplet, including a graviton, a gravitini and a gauge field, and also  $n$ scalar fields and $n$ fermions, which form $\mathcal{N}=2$ scalar multiplet. In \cite{Deger:2002hv,ParkRoLee} it was considered a reduced version of the model  for the sigma-model manifold $H^{2}$
\be\label{sigmaM}
H^2 =SU(1,1)/U(1),
\ee
which corresponds to a single real scalar and other fields are set to zero.
In this case,  the gravity action  is given  by
\be\label{GravAct}
S = \frac{1}{16\pi G_{3}} \int d^{3}x\sqrt{|g|}\left(R -\frac{1}{a^2}(\partial\phi)^2 -V(\phi)\right) + G.H.Y.,
\ee
where $G.H.Y.$ is Gibbons-Hawking-York term and the potential of the scalar field $V$ is given by the following relation:
\be\label{DilPot}
V(\phi)=2\Lambda_{uv}\cosh^2\phi\left[(1-2a^2)\cosh^2\phi+2a^2\right],
\ee
where $\Lambda_{uv}<0$ is a cosmological constant, $a$ is a constant related to a curvature of the sigma-model manifold (\ref{sigmaM}).

\begin{figure}[h!]
	\centering
	\includegraphics[width= 7.5 cm]{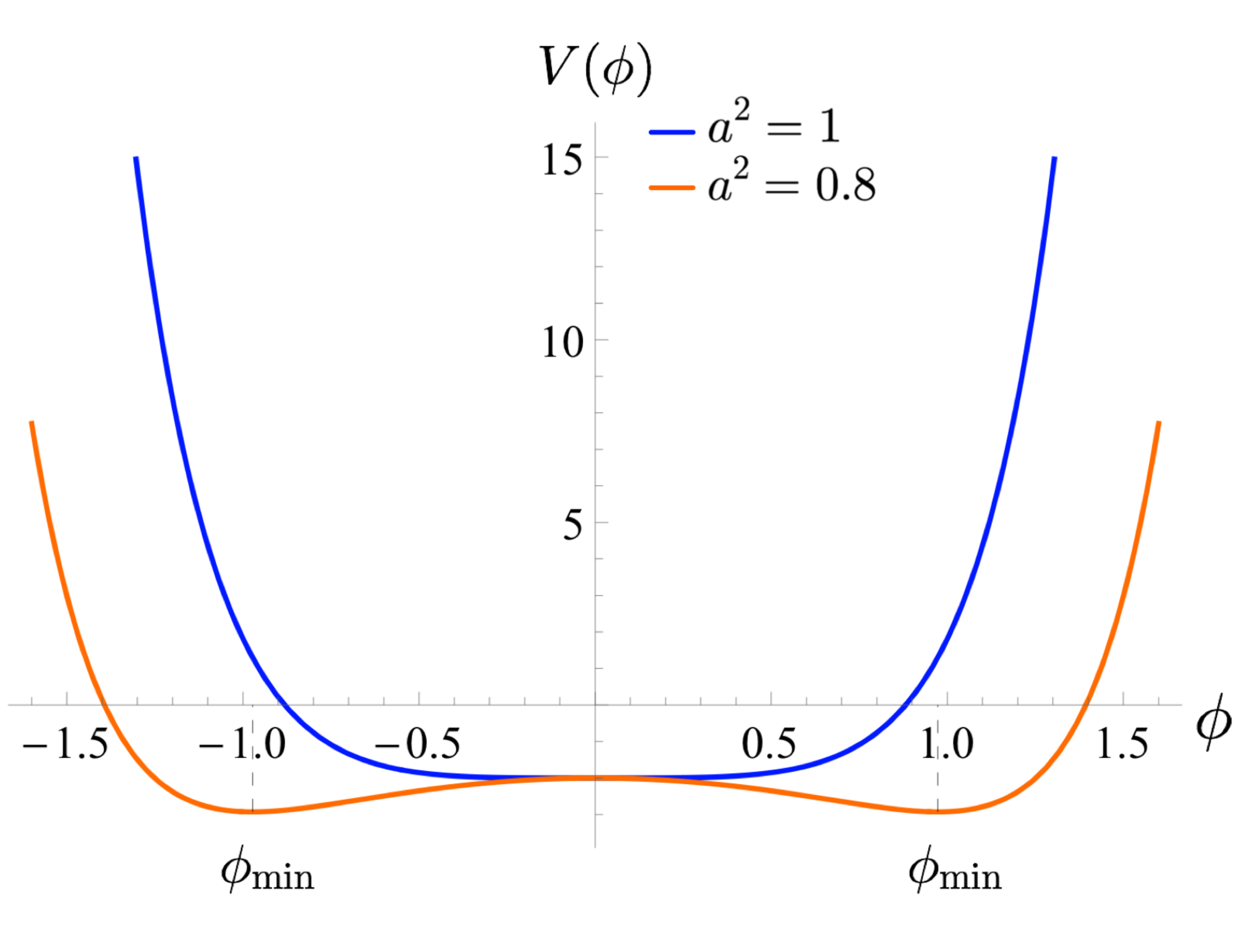}{\bf a)}
	\includegraphics[width= 7.5 cm]{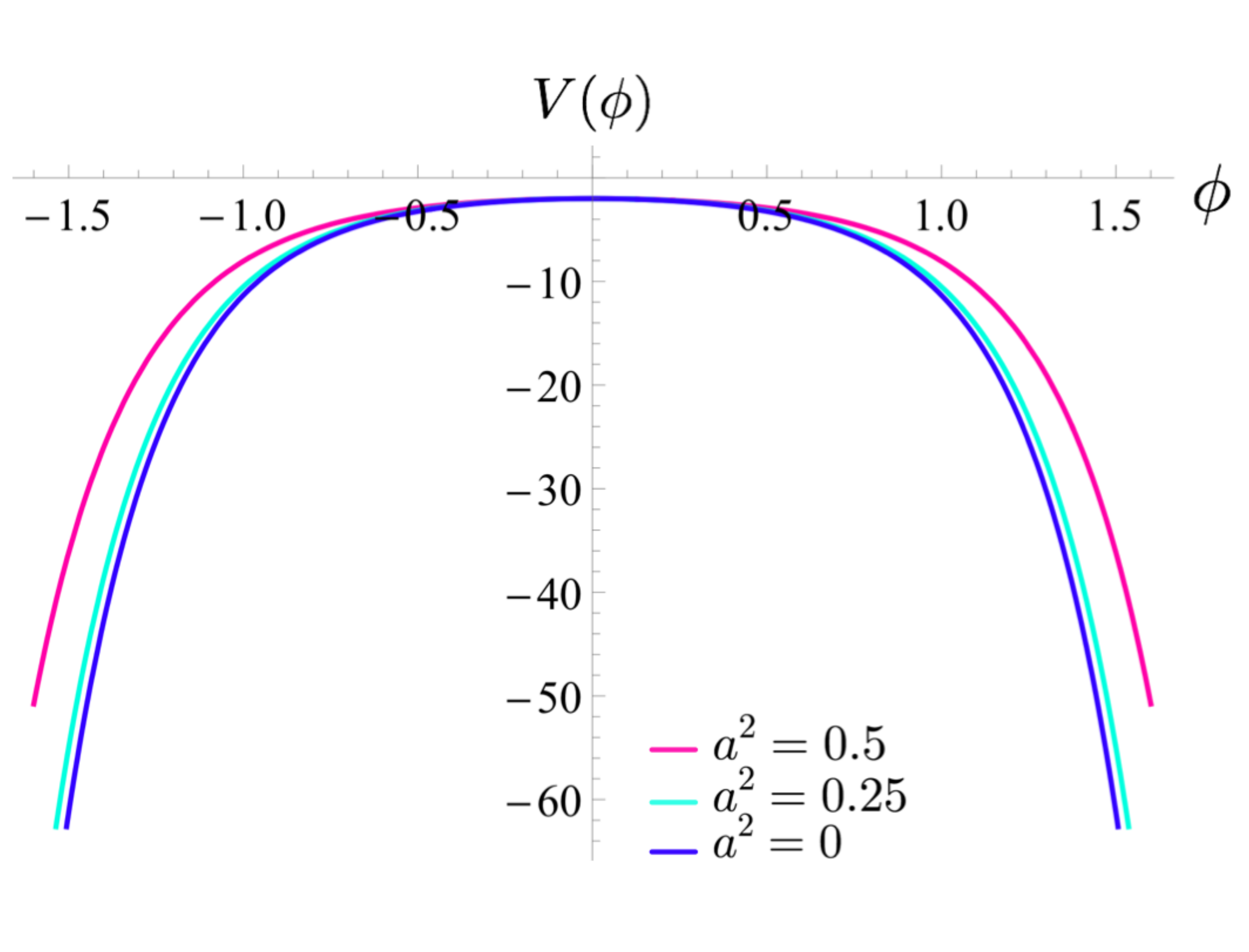}{\bf b)}
	\caption{\small The dependence of the potential $V$ (\ref{DilPot})  on the parameter $a^2$: {\bf a)} the blue line corresponds to the potential at $a^2$=1, the orange line is for $a^2=0.8$; {\bf b)} the pink line represents the behavior of the potential  at $a^2=0.5$, the light blue line is for $a^2=0.25$, the blue line is for $a^2=0$.}
	\label{Fig:DilPot}
\end{figure}
 
 The behavior of the dilaton potential (\ref{DilPot}) depends on $a$.  We show the potential for different values of  $a^2$ on Fig.~\ref{Fig:DilPot}.  It is supposed that the holographic dual theory  (\ref{GravAct}) is some 2d quantum theory with conformal fixed points corresponding to the extrema of the potential.
 
 Note, that the potential (\ref{DilPot}) always has a local extremum $V(0)=2\Lambda_{uv}$ at $\phi=0$. At this point, a  solution of the equations of motion is an $AdS_3$ metric.
 
Let us estimate the behavior of the derivative of the dilaton potential
\be
\frac{dV}{d\phi} = \Lambda_{uv} \left(\left(1-2 a^2\right) \sinh (4 \phi )+2 \sinh (2 \phi)\right).
\ee 
So we have the following extreme points of the potential:
 \be\label{ScFConst}
 \phi_{1} =0, \quad \phi_{2,3} = \frac{1}{2}\ln\left(\frac{1\pm 2|a|\sqrt{1-a^{2}}}{2a^2-1}\right).
 \ee
It worth to be noted that the vacuum, which corresponds to  $\phi_{1}=0$, preserves supersymmetry  both  for $a^2\leq\frac{1}{2}$ and  $a^2>\frac{1}{2}$, while the vacua at  $\phi_{2,3} = \frac{1}{2}\ln\left(\frac{1\pm |a|\sqrt{1-a^{2}}}{2a^2-1}\right)$ are not supersymmetric.
 The corresponding  solutions for the metric are  $AdS_3$ spacetimes. Note, that for $a^2>1/2$ the potential is not bounded above.

 It's also useful to know the behavior of the superpotential 
 \be\label{SuperP}
 W=\sqrt{-\Lambda_{uv}}\cosh^2\phi,
 \ee
 which is related to the potential by the following relation:
 \be
 V(\phi) = \frac{a^2}{4}\left(\frac{\partial W}{\partial \phi}\right)^2 - \frac{1}{2}W^2.
 \ee 
 \begin{figure}[h!]
	\centering
	\includegraphics[width= 7.5 cm]{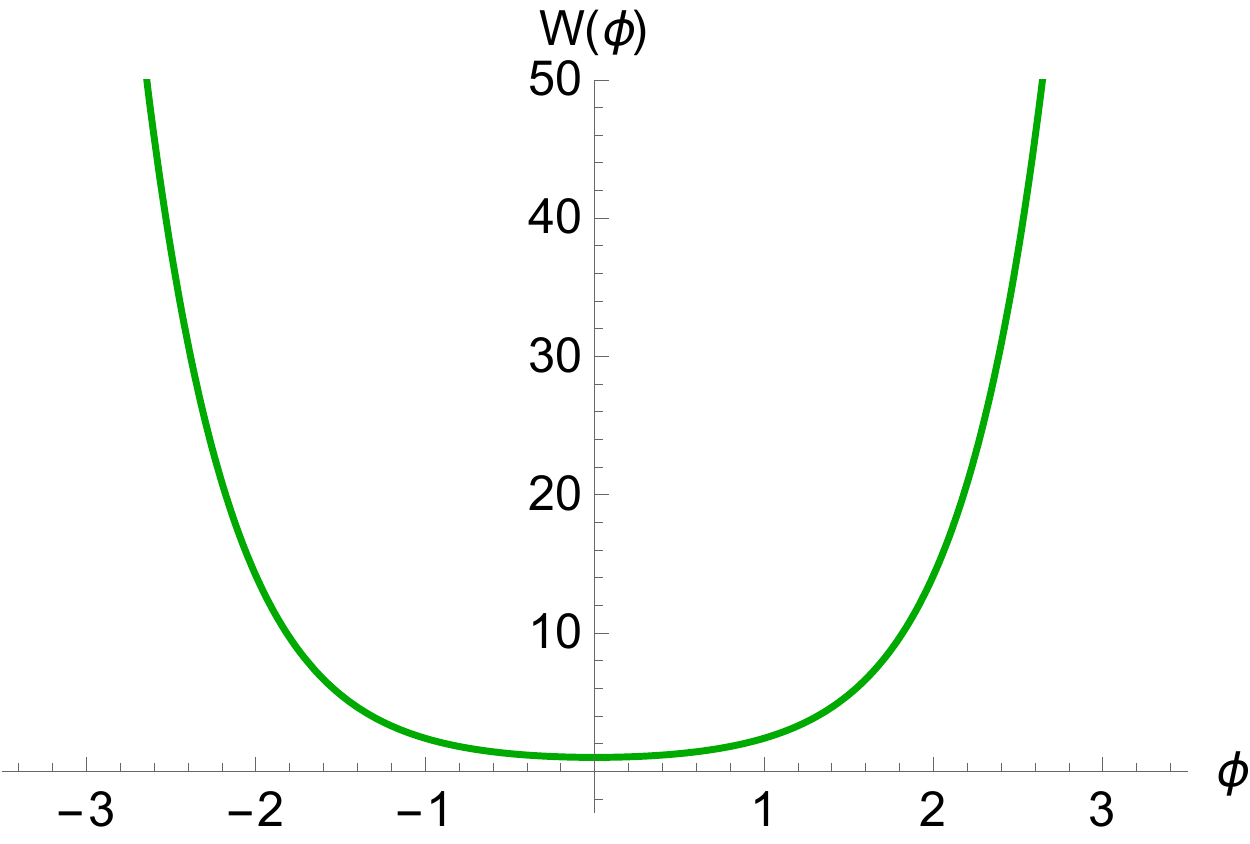}
	\caption{The behavior of the superpotential for the model (\ref{SuperP}).}
	\label{Fig:SuperP}
	\end{figure}

In Fig.~\ref{Fig:SuperP} we plot the dependence of the superpotential $W$ (\ref{SuperP}) on $\phi$. It worth to be noted that
for holographic RG flows the superpotential always increases, thus its minimum corresponds to a UV fixed point, while the maximum is associated with an IR \cite{FGPW}. 

We will look for asymptotic solutions for the metric in domain wall coordinates:
\be\label{metricT}
ds^2=e^{2A(w)}\left(-dt^2+dx^2\right)+dw^2,
\ee
where $w$ is a radial coordinate, which is defined on the interval  $(w_{0}, +\infty)$. We also suppose the following dependence for the scalar field: 
\be
\phi= \phi(w).
\ee
Then the Einstein equations may be written as follows
\bea
2\ddot{A}+2\dot{A}^2+V+\frac{\dot{\phi}^2}{a^2}&=&0,\label{eom1}\\
2\dot{A}^2+V-\frac{\dot{\phi}^2}{a^2}&=&0.\label{eom3}
\eea
 From the equations (\ref{eom1})-(\ref{eom3}) one finds a useful relation
\be\label{eomextra}
\ddot{A}+\frac{\dot{\phi}^2}{a^2}=0.
\ee
The equation for the dilaton has the following form
\be\label{eqd}
\ddot{\phi}+ 2\dot{A}\dot{\phi}- \frac{a^2}{2}V_{\phi}=0.
\ee
Finally, the equations of motion can be represented as follows
\bea
2\dot{A}^2+V-\frac{\dot{\phi}^2}{a^2}&=&0,\label{1T0}\\
\ddot{A}+\frac{\dot{\phi}^2}{a^2}&=&0,\label{3T0}\\
\ddot{\phi}+2\dot{A}\dot{\phi} -  \frac{a^2}{2}V_{\phi}&=&0.\label{4T0}
\eea

Eqs. (\ref{1T0})-(\ref{4T0}) have an exact solution  \cite{Deger:2002hv}. Namely,  the scalar field is represented as follows
	\be\label{scafieldDeg}
	\phi = \frac{1}{2}\ln\left(\frac{1+e^{-4ma^2 w}}{1-e^{-4m a^2 w}}\right), \quad 0\leq w< \infty,
	\ee
	where $m^2=-\frac{\Lambda_{uv}}{4}$.
The scale factor for this case is given by the relation:

	\be\label{scafDeg}
	A = \frac{1}{4a^{2}}\ln (e^{8 m a^2 w}-1).
	\ee
Then the metric (\ref{metricT}) can be represented in the following form
\be\label{metricDeg}
ds^2=(e^{8 m a^2 w}-1)^{\frac{1}{2a^2}}\left(-dt^2+dx^2\right)+dw^2.
\ee
The solution (\ref{scafieldDeg})-(\ref{scafDeg}) preserves half supersymmetry charges, since the constraint for preserving supersymmetries was imposed in the derivation of the first order equations of motion, which relate the derivatives of the scale factor and  the scalar field to the superpotential  and its derivative, correspondingly\cite{Deger:2002hv}.
The superpotential (\ref{SuperP}) with (\ref{scafieldDeg})-(\ref{scafDeg}) takes the form 
\be
W = \frac{2m}{1 - e^{-8ma^2w}}.
\ee
Note that in this work we are focusing on eqs.~(\ref{1T0})-(\ref{4T0}) and supersymmetries are not guaranteed to be preserved.

\setcounter{equation}{0}
 \section{Scaling dimensions of operators} \label{Sect:3}

A holographic RG flow is described by a domain wall solution interpolating between two different asymptotic spacetimes associated to fixed points of a dual field theory. This flow can be related to a deformation of CFT induced in two ways: namely, by a relevant operator or by a non-zero VEV of a scalar operator \cite{BFS}. These quantities can be read off from the asymptotic expansion of the bulk field.

For this, we consider the scalar field in the   $AdS_3$ spacetime
\be\label{Funcd}
S \sim \int d^3 x\sqrt{-g}\left(g^{\mu\nu}(\partial \phi)^2 + m^2 \phi^2 \right),
\ee
where $g$ is the determinant of $AdS_{3}$, $m$ is a mass of the scalar field  $\phi$. It worth to be noted that since the dual field theory is a two-dimensional one, the operators are relevant for $\Delta<2$, marginal for  $\Delta=2$ and irrelevant for  $\Delta>2$.

We expand the dilaton potential (\ref{DilPot}) to a quadratic order near $\phi =0 $, which gives the following relation for the mass:
\be\label{dilatonmass}
m^2=4a^2(a^2-1).
\ee

As already mentioned, the fixed point of the holographic RG flow corresponds to  $AdS_3$. In what follows it is  more convenient to analyze in the domain wall coordinates  (\ref{metricT}), then  the AdS metric is
\be\label{DW-AdS}
ds^2_{DW}=e^{2\sqrt{ -\Lambda_{uv}}(w-w_0)}(-dt^2+dx^2)+dw^2.
\ee
Correspondingly, the equation of motion for the scalar field (\ref{Funcd}) in $AdS_{3}$ using (\ref{DW-AdS}) takes the form:
\be\label{EOMd}
\partial_w^2\phi+ 2\sqrt{-\Lambda_{uv}}\partial_w\phi-m^2\phi=0.
\ee
 We will look for its general solution in the form:
\be\label{Sc-an}
\phi\sim e^{-\Delta(w-w_0)},
\ee
where the parameter $\Delta$ corresponds to the scaling dimension of the operator.
Plugging (\ref{Sc-an}) into  (\ref{EOMd}),  we find the constraint for the scalar field mass  (the Breitenlohner-Freedman bound)
\be\label{BFound2}
\Delta(\Delta-2\sqrt{-\Lambda_{uv}})-m^2=0,
\ee
or
\be
\Delta_{\pm}=\sqrt{-\Lambda_{uv}}\pm\sqrt{-\Lambda_{uv}+m^2}.
\ee
So, the Breitenlohner-Freedman bound (we set $\Lambda_{uv} = -1$) is given by
\be\label{BFbound}
m^2\geq -1.
\ee
Taking into account (\ref{dilatonmass}) we get

\be\label{DeltaOp}
\Delta = \Delta_{+} = 1+|1- 2a^2|.
\ee

From (\ref{BFbound}) with (\ref{dilatonmass}) it follows that the parameter $a$ with respect to the Breitenlohner-Freedman bound takes the following values: 
\be
a^2\in\left[0;\frac{1+\sqrt{2}}{2}\right].
\ee
At the same time, we have the following cases for the scaling dimension depending on $a^2$:
\begin{enumerate}
    \item for $a^2=0$ and $a^2=1$ the only value $\Delta=2$, i.e. the operator is marginal,
    \item  for $a^2\in (0;1/2)\cup(1/2;1)$ the scaling dimension  takes the values in the range  $1<\Delta<2$, so the operator is relevant,
    \item for $a^2=1/2$ the only possible value $\Delta=1$, so the operator is relevant.
    \item for $1<a^2\leq\frac{1+\sqrt{2}}{2}$ the operator is irrelevant, since $\Delta>2$.
\end{enumerate}

One can write a generic solution for the scalar field in the asymptotically AdS background (\ref{metricT}) with the required scale factor $A(w)\sim w$ (as $w\to \infty$) \cite{Papadimitriou:2007sj}:
\be\label{ScalarFieldGen} 
\phi \sim \begin{cases}
\phi_0^{-}e^{-(2-\Delta)w}+ \phi_0^{+} e^{-\Delta w}, & \quad \Delta > 1 \\
e^{-w}w\phi_0^{-} +e^{-w}\phi _{0}^{+},  &\quad \Delta=1.
\end{cases}.
\ee
We can associate solutions behaving as $e^{-(2-\Delta)w}$ with deformations of the dual CFT by some operator $\mathcal{O}_{\phi}$. The solutions, which are asymptotic to
$e^{-\Delta w}$ can be associated with deformations triggered by a non-zero VEV of a scalar operator.

Thus, assuming the Dirichlet boundary conditions, the exact solution (\ref{scafieldDeg})-(\ref{scafDeg}) with $0< a^2<1/2$  can be considered as a deformation by a relevant operator or a marginal operator for $a^2=0$ and  $a^2=1$. For 
$a^2 = \frac{1}{2}$ the solution can be interpreted as a deformation induced  by a non-zero vacuum expectation value  of some operator $\mathcal{O}_{\phi}$. 
For $1/2 <a^2<1$ and $1<a^2\leq\frac{1+\sqrt{2}}{2}$ the solution can be related to a flow triggered by a non-zero VEV  of an operator $\mathcal{O}_{\phi}$. It worth to be noted that the domain wall solution (\ref{scafieldDeg})-(\ref{scafDeg}) is supersymmetric and it is required $\Braket{T_{ij}} =0$ \cite{Bianchi:2001de}, which is satisfied under the Dirichlet boundary conditions \cite{Papadimitriou:2004rz}.
 Note, that the solutions (\ref{scafieldDeg})-(\ref{scafDeg}) with $\frac{1}{2}<a^2<1$ or  $a^2\geq 1$ are unphysical, since the scalar field potential is not bounded from above for this choice of $a^2$.
 
The model (\ref{GravAct})-(\ref{DilPot}) admits $AdS_{3}$ solutions for $\frac{1}{2}<a^2<1$, which are not supersymmetric. One can also linearize the potential (\ref{DilPot}) near extremum (a local minimum) $\cosh\phi^2 = \frac{a^2}{2a^2 -1}$ (\ref{ScFConst}). Then we get another relation for the mass of the scalar field:
\be\label{m2a22}
m^2 = 8 (a^2-1)
\ee
and we get for the  scaling  dimension
\be\label{DeltaOp2}
\Delta = \Delta_{+}= 1+\sqrt{9-8a^2}.
\ee
So the operator tends to be irrelevant with $\frac{1}{2}<a^2<1$. 

In \cite{Papadimitriou:2007sj} to study multi-trace deformations in the context of the holographic duality it was suggested to impose  Neumann or the Mixed boundary conditions on the bulk fields. 
In  this case the BF bound is
\be\label{BFbound2}
-1\leq m^2\leq 0.
\ee
Taking into account (\ref{dilatonmass}) and (\ref{m2a22})  we get the constraints for the parameter $a^2$ 
\be
 \begin{cases}
1\leq a^2\leq\frac{1+\sqrt{2}}{2}, & \quad \phi =0, \\
\frac{7}{8}\leq a^2 \leq 1,  &\quad \phi_{2,3} = \frac{1}{2}\ln\left(\frac{1\pm 2|a|\sqrt{1-a^{2}}}{2a^2-1}\right).
\end{cases}
\ee
As it was mentioned  the supersymmetric solution with  $a^2>1$ is unphysical because of a naked singularity,
while  a non-supersymmetric kink solution between two AdS vacua with Neumann/the Mixed boundary conditions is of interest. In this case the kink solution can be associated  with a multi-trace deformation  \cite{Papadimitriou:2007sj}.

\setcounter{equation}{0}
\section{Dynamical system}

In this section, we study the holographic renormalization group flow by means of the dynamical system theory.
The equations of motion (\ref{1T0})-(\ref{4T0}) are reduced to an autonomous dynamical system, that allow to classify the stability of stationary (critical) points.
According to the holographic approach, these points correspond to the fixed points of the dual theory.

\subsection{Autonomous dynamical system}

To obtain a system of differential equations describing the RG flow, we introduce a new variable \cite{KLN}-\cite{IAHoReGR}
\be\label{XvarDef}
X=\frac{\dot{\phi}}{\dot{A}}
\ee
where  $\phi$ is the dilaton,  $A$ is the scale factor, and the derivatives are taken with respect to the radial variable $w$.
It is also convenient to introduce the following variable
\be\label{Zvar}
Z = e^{-\phi},
\ee
so that $Z\in(0, +\infty)$ for $\phi \in(-\infty;\infty)$. Note, that such a replacement will not allow to find and display points with $Z\in \infty$ on the phase diagrams.  For these fixed points, the asymptotic behavior of the dilaton looks like $\phi\to -\infty$, and it corresponds to a small value of the coupling constant of the dual theory, thus we omit this case.

Using (\ref{Zvar}) the scalar field can be expressed as 
\be
\phi= -\ln Z\label{03}.
\ee
The variables $X$ and $Z$ allow to rewrite the second-order differential equations (\ref{1T0})-(\ref{4T0}) as a dynamical system of first-order differential equations.
 Thus, the study of holographic RG flows can be implemented in terms of the new variables $X$ and $Z$.

Dividing eq. (\ref{1T0}) by ${\dot A}^2$ and doing some algebra we obtain
\be\label{pr1}
\frac{1}{\dot{A}^2}=\frac{1}{V}\left(\frac{X^2}{a^2}-2\right).
\ee
For the equation (\ref{3T0}) we have
\be\label{pr2}
\frac{\ddot{A}}{\dot{A}^2}=-\frac{X^2}{a^2}.
\ee
Finally, we get the following expression for (\ref{4T0}) after deriving it by $\dot{A}^2$ with respect to (\ref{pr1})-(\ref{pr2})
\be\label{pr3}
\frac{\ddot{\phi}}{\dot{A}^2}=-2X+ \frac{a^2}{2}\frac{V_{\phi}}{V}\left(\frac{X^2}{a^2}-2\right).
\ee
The potential and its derivative are rewritten in terms of the variable $Z$
\bea\label{V}
V&=&\frac{\Lambda  \left((Z^2+1)^4-2 a^2 (Z^4-1)^2\right)}{8 Z^4},\\
V_{\phi} &=&\frac{\Lambda  \left(2 a^2 (Z^8-1)-(Z^2-1) (Z^2+1)^3\right)}{2 Z^4},
\eea
so we have the following ratio
\be\label{VphV}
\frac{V_{\phi}}{V}= \frac{4\left(2a^2(Z^8 - 1)-(Z^2 -1)(Z^2+1)^3\right)}{(Z^2 +1)^4 - 2a^2(Z^4 - 1)^2}.
\ee
To obtain the system of first-order differential equations, we note that $\frac{d}{dA}=\frac{d}{\dot{A}dw}$, then we will obtain $\frac{dZ}{dA}$, $\frac{dX}{dA}$ as functions of the variables $Z$, $X$. 
In this case the system of first-order differential equations, which is equivalent to the system (\ref{1T0})-(\ref{4T0}), may be represented as follows
\bea
\frac{dZ}{dA}=f(Z,X),\label{sis1}\\
\frac{dX}{dA}=g(Z,X),\label{sis2}
\eea
where the functions $f$ and $g$  are defined as
\bea\label{sysLHS}
f(Z,X)&=&-ZX,\\  \label{sysLHS2}
g(Z,X)&=&\left(\frac{X^2}{a^2}-2\right)\Bigl(X+\frac{a^2}{2}\times\frac{V_{\phi}}{V}\Bigr),
\eea
and the ratio $\frac{V_{\phi}}{V}$ is given by (\ref{VphV}).
The equilibrium points of the dynamical system (\ref{sis1})-(\ref{sysLHS2}), corresponding to the fixed points of the field theory, may be found as the result of equality the left part of the system  (\ref{sis1})-(\ref{sis2}) to zero. 

Thus, we have
\bea\label{DynSysY0}
\begin{cases}
	f(Z,X)\bigg|_{Z_c,X_c}=0,\\
	g(Z,X)\bigg|_{Z_c,X_c}=0,
\end{cases}
\eea
or, equivalently,
\bea\label{din}
\begin{cases}
	ZX=0,\\
	\left(\frac{X^2}{a^2}-2\right)\Bigl(X+a^2\times \frac{2\left(2a^2(Z^8 - 1)-(Z^2 -1)(Z^2+1)^3\right)}{(Z^2 +1)^4 - 2a^2(Z^4 - 1)^2}\Bigr) =0. 
\end{cases}
\eea
 
Eqs. (\ref{din}) have the following solutions, which are  the equilibrium points: 
 \begin{enumerate}
 \item $Z_c=0$,  $X_c = - a\sqrt{2}$, 
  \item $Z_c=0$,  $X_c = a\sqrt{2}$,
 \item $Z_c = 0$,  $X_c = -2a^2$,
 \item $Z_c =1$, $X_c =0$, 
 \item $Z_c= \sqrt{\frac{1- 2|a|\sqrt{1 -a^2}}{2a^2 -1}}$, $X_c=0$, 
 \item $Z_c= \sqrt{\frac{1+ 2\sqrt{1 -a^2}}{2a^2 -1}}$, $X_c=0$.
 \end{enumerate}
 Not all of these vacua are supersymmetric, since the dynamical system  (\ref{sis1})-(\ref{sysLHS2}) is obtained from (\ref{1T0})-(\ref{4T0}) without constraints on preserving supersymmetries.
 It should be noted that the equilibrium points with $Z_c=0$ correspond to asymptotic solutions  with non-trivial scalar fields and non-anti-de Sitter metrics, while constant scalar fields and an $AdS_3$ metrics are related to some constant value of
 $Z_c$. We will discuss this below.

\subsection{Stability analysis of equilibrium points}

To classify the equilibrium points, we make linear perturbation $\delta Z$, $\delta X$ near each critical point $(Z_c, X_c)$: 
\be
Z =Z_c +\delta Z,\quad X =X_c +\delta X.
\ee
 The linearization of the dynamical system (\ref{din}) leads to equations in the following form
\begin{equation}
\frac{d}{dA}
\begin{pmatrix}
\delta Z \\
\delta X
\end{pmatrix}
= \mathcal{M}
\begin{pmatrix}
\delta Z \\
\delta X
\end{pmatrix},
\end{equation}
where $\mathcal{M}$ is the Jacobian matrix
\be
\mathcal{M}=\begin{pmatrix} 
\frac{\partial f}{\partial Z} & \frac{\partial f}{\partial X}\\ 
\frac{\partial g}{\partial Z} & \frac{\partial g}{\partial X} 
\end{pmatrix}\Big{\vert}_{Z=Z_c, X=X_c}
\ee
with elements: 
\bea
\mathcal{M}_{11} &=& -X_c, \quad \mathcal{M}_{12}= -Z_c,\\
 \mathcal{M}_{21}&=&-\frac{8 Z_c (2 a^2-X^2_c) \left(8 a^4 Z^2_c (Z^2_c-1)^2+2 a^2 (Z^2_c+1)^2 (Z^4_c+1)-(Z^2_c+1)^4\right)}{(Z^2_c+1)^2 \left((Z^2_c+1)^2-2 a^2 (Z^2_c-1)^2\right)^2},\nonumber\\
 \,  \\
 \mathcal{M}_{22}&=& \frac{3 X^2_c}{a^2}-2 - \frac{4 X_c \left((Z^2_c-1)
   (Z^2_c+1)^3 - 2 a^2 (Z^8_c-1)\right)}{(Z^2_c + 1)^4-2 a^2 (Z^4_c - 1)^2}.
\eea
Then  the characteristic equation takes the form
\be
\lambda^2-\lambda\left(\mathcal{M}_{11}+\mathcal{M}_{22}\right)+\mathcal{M}_{11}\mathcal{M}_{22}-\mathcal{M}_{12}\mathcal{M}_{21}=0.
\ee
Let's write out the determinant of the Jacobian matrix $\det{\mathcal{M}}$ and eigenvalues for each equilibrium point.

{\bf 1.}  For $Z_c=0$, $X_c=a\sqrt{2}$:
\be
\lambda_{1}=-a\sqrt{2};\quad\lambda_{2}=4\left(1+a\sqrt{2}\right)
\ee
and
\be
\det{\mathcal{M}}=-4\sqrt{2}a(1+\sqrt{2}a).
\ee
The equilibrium point is an unstable node for $a\in(-\frac{1}{\sqrt{2}};0)$, while  for $a\in(0;\frac{1}{\sqrt{2}}]$  and $a^2\in(\frac{1}{2};1]$it is a saddle.
Note, the stability of this equilibrium point is not determined for $a=0$ and $a=-\frac{1}{\sqrt{2}}$. 

 {\bf 2.} For $Z_c=0$, $X_c=-a\sqrt{2}$:
\be
\lambda_{1} =\sqrt{2}a,\quad \lambda_2 = 4(1-a\sqrt{2})
\ee
and
\be
\det{\mathcal{M}} = 4\sqrt{2}a(1-\sqrt{2}a).
\ee
Note, as for the previous case, the stability of the critical point depends on the sign of the parameter $a$ when $a^2\in (0;\frac{1}{2})$. So, for the values $a\in(0;\frac{1}{\sqrt{2}})$ it is an unstable node and for  $a\in[-1;0)\cup(\frac{1}{\sqrt{2}};1]$ and it is a saddle.
The stability of the equilibrium point is undefined for $a=0$ and $a=\frac{1}{\sqrt{2}}$.

{\bf 3.} For $Z_c=0$, $X_c= -2a^2$:
\be\label{eigenC3}
\lambda_{1}=2a^2,\quad\lambda_{2}=2(2a^2 -1)
\ee
and
\be
\det{\mathcal{M}}=4a^2(2a^2-1).
\ee

  We are not able to identify the type of stability of this equilibrium point for $a^2=0$ and $a^2=\frac{1}{2}$ by means of a linear analysis. For $a^2\in(0;\frac{1}{2})$ it is a saddle, for  $a^2\in(\frac{1}{2};1]$ it is an unstable node. 
  
{\bf 4.} For $Z_c=1$, $X_c=0$:
\be
\lambda_{1}=-2a^2,\quad\lambda_{2}=2(a^2 -1)
\ee
and
\be
\det{\mathcal{M}}=4a^2(1-a^2).
\ee

The critical point is a stable node for $a^2\in(0;1)$, however its type of stability cannot be identified by means of the linear perturbations for two cases: $a^2=0$ and $a^2=1$.\\

To discuss the points {\bf 5} and {\bf 6} we note,
 that for $X_c = 0$ and  an arbitrary $Z_c$, the elements of the Jacobian matrix have the form:
\be
\mathcal{M}_{21}= \frac{16 a^2 Z_c (8 a^4 Z^2_c (Z^2_c-1)^2+2 a^2 (Z^2_c+1)^2
   (Z^4_c+1)-(Z^2_c+1)^4)}{(Z^2_c+1)^2((Z^2_c+1)^2-2 a^2 (Z^2_c-1)^2)^2}, 
   \ee
  where
   \be
   Z_c   = \sqrt{\frac{1\pm 2|a|\sqrt{1-a^2}}{2a^{2}-1}}.
   \ee 
  For both cases, we focus on $a^{2}> \frac{1}{2}$, since  we get a complex $Z_c$ for $a^2\in (0;\frac{1}{2})\cup(1;+\infty)$, that yields a complex scalar field.\\

   {\bf 5.} For $Z_c=\sqrt{\frac{1- 2|a|\sqrt{1-a^2}}{2a^{2}-1}}$, $X_c=0$  with $a^{2}> \frac{1}{2}$, the following eigenvalues are obtained: 
   \be
   \lambda_{1,2}=-1\pm\sqrt{9- 8a^2 },
   \ee
    the determinant of the Jacobian matrix  reads:
\be
\det{\mathcal{M}} =8(a^2-1).
\ee

The critical point is a saddle for $a^2\in (\frac{1}{2};1)$ and its stability is unknown for $a^2=1$.

{\bf 6.} For  $Z_c=\sqrt{\frac{1+ 2|a|\sqrt{1-a^2}}{2a^{2}-1}}$, $X_c=0$  for $a^{2}> \frac{1}{2}$, we get
   \be
   \lambda_{1,2}=-1\pm\sqrt{9- 8 a^2 }.
   \ee
 The determinant of the Jacobian matrix is
\be
\det{\mathcal{M}}=8( a^2-1).
\ee

As well as in the previous case, the critical point is a saddle for $a^2\in (\frac{1}{2};1)$, while the of stability is unknown for the parameter value $a^2=1$.

In the table \ref{TableTwo} we summarize the results of our stability analysis for different values of the parameter $a$.\\
\begin{table}
\scriptsize
\noindent	{\extrarowheight = 2pt
	\begin{tabular}{|c|l!{\vrule width 0.1pt\relax}c|l!{\vrule width 0.3pt\relax}c|l!{\vrule width 0.1pt\relax}c|}
		\hline
		\multirow{2}{*}{point} & \multicolumn{2}{c|}{$a^2=0$} &\multicolumn{2}{c|}{$0<a^2< \frac{1}{2}$}&\multicolumn{2}{c|}{$a^2=\frac{1}{2}$}\\
		\cline{2-7}
		& \multicolumn{2}{c|}{$V =2\Lambda_{uv}\cosh^4\phi $} & \multicolumn{4}{c|}{$V = 2\Lambda_{uv}\cosh^2\phi\left[(1-2a^2)\cosh^2\phi+2a^2\right]$}\\
		\hline
		1 & \begin{tabular}{>{$}c<{$}}
			\lambda_1 = 0  \\
			\lambda_2 = 4
		\end{tabular} & \multirow{2}{*}{none} &${\bf -\frac{1}{\sqrt{2}}<a<0}:\quad\begin{matrix}
0<\lambda_1<1\\
0<\lambda_2<4
\end{matrix}$ & unst.node&${\bf a=\frac{1}{\sqrt{2}}:}\quad
\begin{matrix}
\lambda_1=-1\\
\lambda_2=8
\end{matrix}$ & saddle \\
		\cline{4-7}& & & ${\bf 0<a<\frac{1}{\sqrt{2}}:}\quad
		\begin{matrix}
		-1<\lambda_1<0\\
		4<\lambda_2\leq8 
		\end{matrix}$ & saddle&${\bf a=-\frac{1}{\sqrt{2}}:}\quad
		\begin{matrix}\lambda_1=1\\
		\lambda_2=0
		\end{matrix}$ & none\\
		\hline
		\multirow{2}{*}{2} &	\multirow{2}{*}{\begin{tabular}{>{$}c<{$}}
				\lambda_1 = 0 \\
				\lambda_2 = 4
		\end{tabular} } & 	\multirow{2}{*}{none} &${\bf 0<a<\frac{1}{\sqrt{2}}:}\quad
		\begin{matrix}
		0<\lambda_1< 1\\ 0<\lambda_2<4
		\end{matrix}$ & unst.node&${\bf a=\frac{1}{\sqrt{2}}:}\quad
		\begin{matrix}
		\lambda_1=1\\
		\lambda_2=0
		\end{matrix}$  &none\\
		\cline{4-7}& & & ${\bf -\frac{1}{\sqrt{2}}<a<0:}\quad
		\begin{matrix}
		-1\leq\lambda_1< 0\\ 4<\lambda_2\leq 8
		\end{matrix}$ & saddle&  ${\bf a=-\frac{1}{\sqrt{2}}:}\quad
		\begin{matrix}
		\lambda_1=-1\\
		\lambda_2=8
		\end{matrix}$&saddle\\
		\hline
		3 &  \begin{tabular}{>{$}c<{$}}
			\lambda_1 = 0  \\
			\lambda_2 = -2 
		\end{tabular}  & none & 
		$	0<\lambda_1<1,\quad-2<\lambda_2<0 $
		& saddle&${\bf  a^2=\frac{1}{2}}:\quad \lambda_1=1,\quad\lambda_2=0$ & none\\	
		\hline
		4 &  \begin{tabular}{>{$}c<{$}}
			\lambda_1 = 0  \\
			\lambda_2= -2 
		\end{tabular}  & none & 
		\begin{tabular}{>{$}c<{$}}
			-1<\lambda_1 <0  \\
			-2<	\lambda_2 <-1
		\end{tabular} & stab.node&${\bf  a^2=\frac{1}{2}}:\quad \lambda_{1,2}=-1$& stab.node\\	
		\hline
		5,6 &  \begin{tabular}{>{$}c<{$}}
			\lambda_1 =2  \\
			\lambda_2 = -4
		\end{tabular}  & saddle & 
		\begin{tabular}{>{$}c<{$}}
			2<\lambda_1<-1+\sqrt{5}  \\
			-4<	\lambda_2 <-1-\sqrt{5}
		\end{tabular} & saddle&${\bf  a^2=\frac{1}{2}}:\quad \lambda_{1,2}=-1\pm\sqrt{5}$ &saddle \\	
		\hline
		
	\end{tabular}
}

\noindent	{	\extrarowheight = 2pt \scriptsize
	\begin{tabular}{|c|l!{\vrule width 0.1pt\relax}c|l!{\vrule width 0.3pt\relax}c|}
		\hline
		\multirow{2}{*}{point} & \multicolumn{2}{c|}{$\frac{1}{2}<a^2<1$}&\multicolumn{2}{c|}{$a^2=1$} \\
		\cline{2-5}&
		 \multicolumn{2}{c|}{$V = 2\Lambda_{uv}\cosh^2\phi\left[(1-2a^2)\cosh^2\phi+2a^2\right]$}& \multicolumn{2}{c|}{$V = 2\Lambda_{uv}\cosh^2\phi\left[-\cosh^2\phi+2\right]$} \\
		\hline
		\multirow{2}{*}{1} &  
		${\bf a<0}:
		\begin{matrix}
		1<\lambda_1<\sqrt{2}\\ 4(1-\sqrt{2})<\lambda_2<0 
		\end{matrix}$		
	&\multirow{2}{*}{saddle} &
			${\bf a<0 }:\quad
			\begin{matrix}
			\lambda_1=\sqrt{2}\\
			\lambda_2=4(1-\sqrt{2})
			\end{matrix}$
	& \multirow{2}{*}{saddle} \\	
	\cline{2-2}\cline{4-4}
	& ${\bf  a > 0}:\quad
		\begin{matrix}-\sqrt{2}<\lambda_1<-1\\ 8<\lambda_2<4(1+\sqrt{2})
			\end{matrix}$
& &$ {\bf a>0}: \quad
	\begin{matrix}\lambda_1=-\sqrt{2} \\	\lambda_2=4(1+\sqrt{2})
		\end{matrix}$& \\
		\hline
		\multirow{2}{*}{2} &  
			${\bf \frac{1}{\sqrt{2}}<a<1:}\quad
				\begin{matrix}
				1<\lambda_1<\sqrt{2}\\
				4(1-\sqrt{2}) <\lambda_2<0
					\end{matrix}$ & \multirow{2}{*}{saddle} &
		$\text{I}) \quad
		\begin{matrix}\lambda_1 =\sqrt{2}\\
		\lambda_2=4(1-\sqrt{2}) 	\end{matrix}$ 		 
	 &\multirow{2}{*}{ saddle}\\	
		\cline{2-2}\cline{4-4} 
		& ${\bf -1<a<-\frac{1}{\sqrt{2}}:}\quad 
		\begin{matrix}
		-\sqrt{2}<\lambda_1<-1\\	8<\lambda_2<4(1+\sqrt{2}) 	\end{matrix}$& &$ \text{II}) \quad		\begin{matrix}\lambda_2 = -\sqrt{2}\\ \lambda_2=4(1+\sqrt{2}) 
			\end{matrix}$  & \\
		\hline 
		3 & 
		\begin{tabular}{>{$}c<{$}}
			1<\lambda_1 <2,\quad
			0<\lambda_2 <2 
		\end{tabular} & unst.node &
			$\lambda_{1,2} = 2$ & unst.node \\	
		\hline
			4 &  
		\begin{tabular}{>{$}c<{$}}
	-2<	\lambda_1<-1  \\
		-1<\lambda_2<0 
		\end{tabular} & st.node &
		\begin{tabular}{>{$}c<{$}}
			\lambda_1 = -2  \\
			\lambda_2 = 0
		\end{tabular} & none \\	
		\hline
			5,6 &  
		\begin{tabular}{>{$}c<{$}}
		0<\lambda_1 <-1+\sqrt{5}  \\
		-1-\sqrt{5}	<\lambda_2 < -2
		\end{tabular} & saddle &
		\begin{tabular}{>{$}c<{$}}
			\lambda_1 = 0  \\
			\lambda_2 = -2 
		\end{tabular} & none \\	
		\hline
		
	\end{tabular}
}
\caption{\small Classification of the equilibrium points of the system (\ref{sis1})-(\ref{sysLHS2}) for different $a$.}
\label{TableTwo}
\end{table}
\normalsize

The phase portraits of the dynamical system (\ref{sis1})-(\ref{sysLHS2}) are presented in Figs. 3, 4, 5 and 6, correspondingly, for $a^2 =\{0.25, 0.5, 0.8, 1\}$. 
For all plots, the abscissa axis
is $Z$, the ordinate axis is $X$. 

For $a^2 =0.25$ ({\bf  Fig.3})
we see four equilibrium points on the phase portrait: the point {\bf 1}  with coordinates $Z_c=0, X_{c}=\frac{1}{\sqrt{2}}$, the point {\bf 2}  with $Z_c=0, X_{c}=-\frac{1}{\sqrt{2}}$, the point {\bf 3} with $Z_c=0,X_c=\frac{1}{2}$ and at last, the point  {\bf 4} with $Z_c=1, X_c=0$. Note that the points {\bf 5} and  {\bf 6} are missing on the phase portrait, since for {\bf 5} we have  $Z_c=1$, so the point coincides with the point {\bf 4}, and for the point {\bf 6} $Z_c$-coordinate is complex.

 For $a^2 =0.5$ ({\bf  Fig.4}) we observe only three points. Points {\bf 1} and {\bf 3} have the same coordinates $Z_c=0,X_c=1$, so, they merged.   Points {\bf 2} and {\bf 4} have coordinates  $Z_c=0,X_c=-1$ and, correspondingly,  $Z_c=1,X_c=0$.  There are also no points {\bf 5} and {\bf 6}, as for $a^2 =0.25$, since $Z_c $ tends to be infinite in these cases.

\begin{figure}[h!]
	\centering
	\includegraphics[width= 7.55cm]{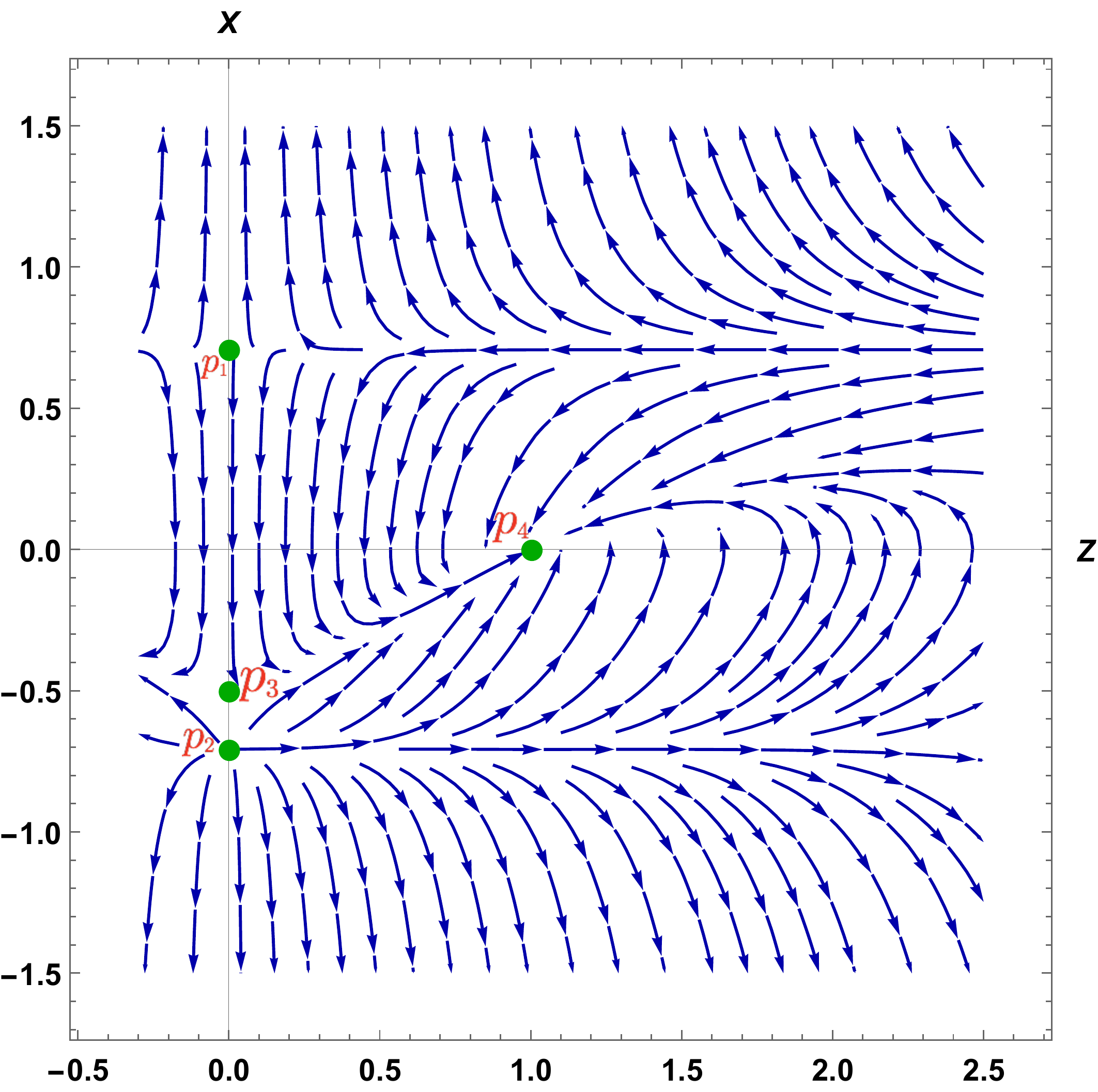}{\bf a)}
	\includegraphics[width= 7.55cm]{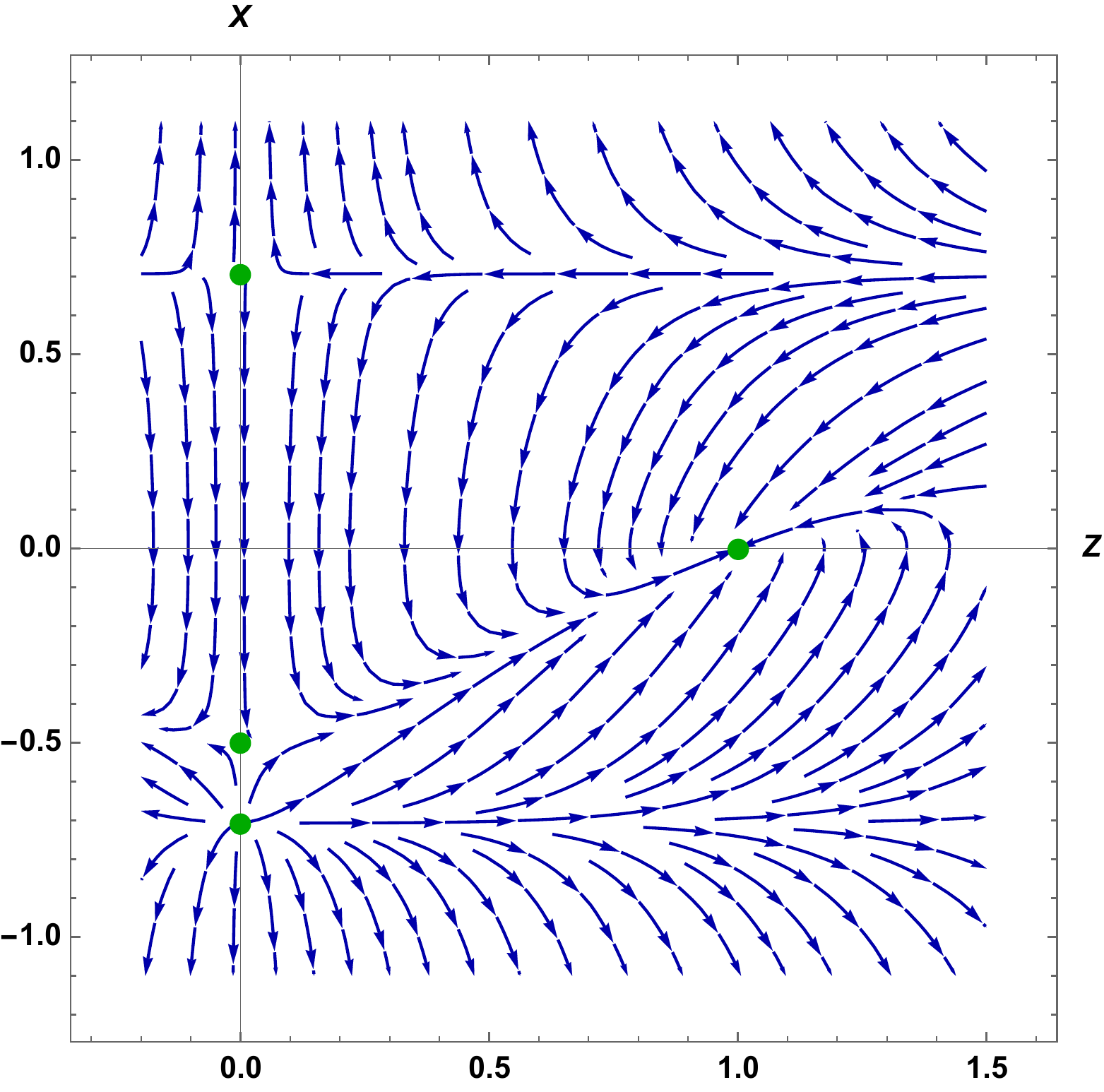}{\bf b)}
	\caption{\small {\bf a)} The phase portrait of the dynamical system (\ref{sis1})-(\ref{sysLHS2}) for $a^2=0.25$ with the exact solution (orange line); {\bf b)} The phase portrait fragment on the interval $Z\in\left[-0.1, 1.5\right]$ and $X\in\left[-1.1, 1.1\right]$.}
	\label{fig:Fugure3}
\end{figure}

	\begin{figure}[h!]
	\centering
	\includegraphics[width= 7.55cm]{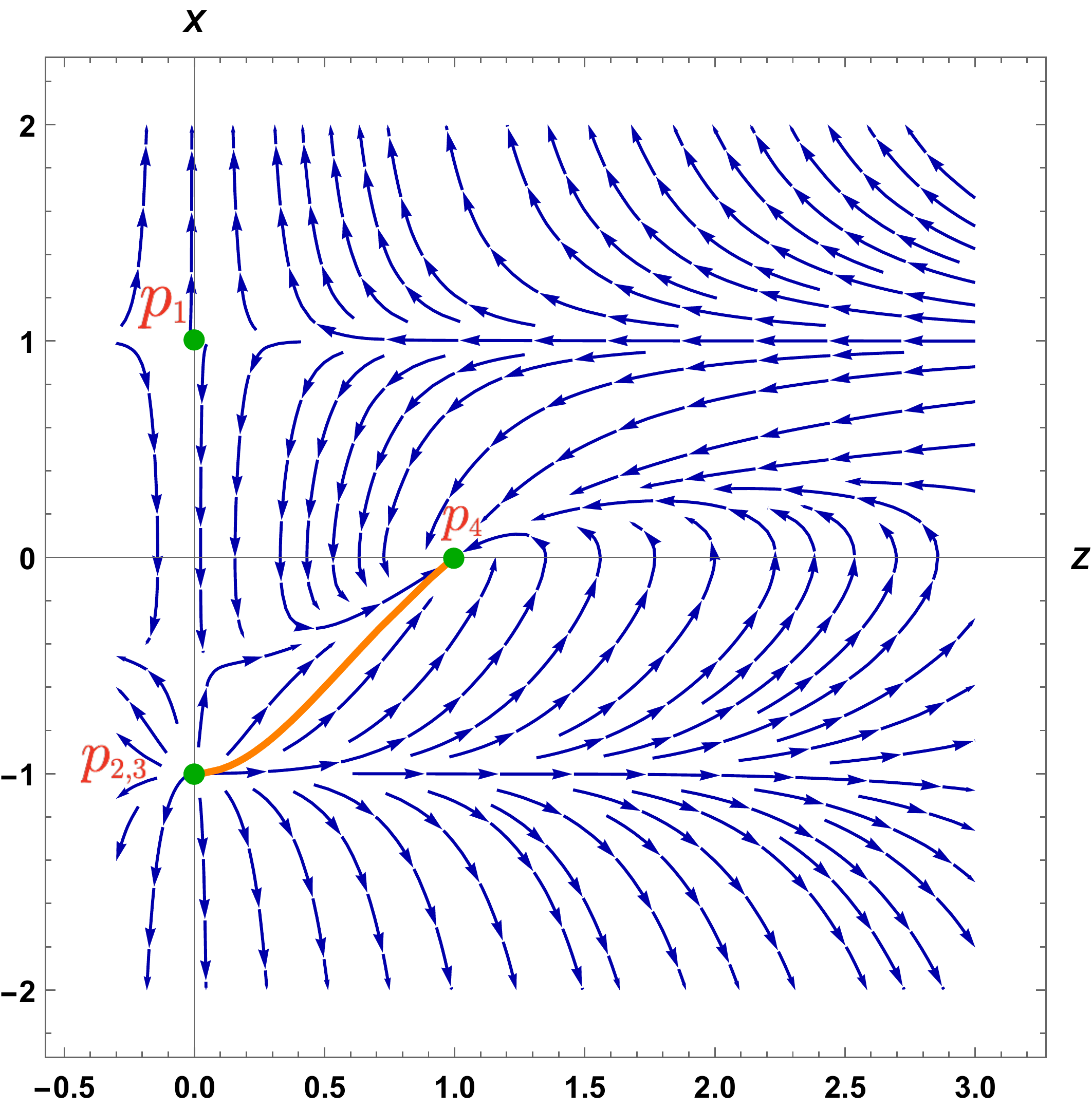}{\bf a)}
	\includegraphics[width= 7.55cm]{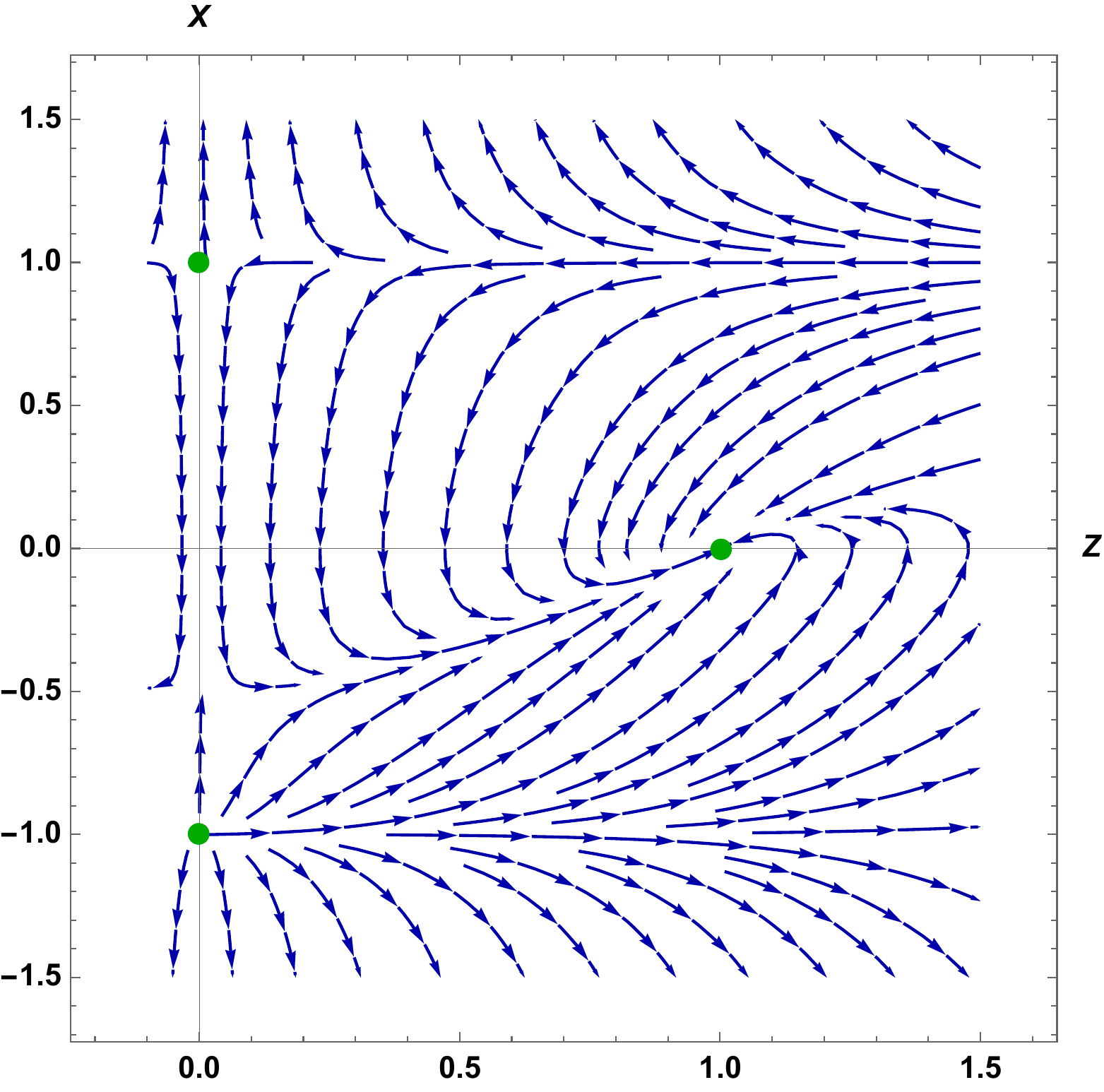}{\bf b)}
	\caption{\small {\bf a)} The phase portrait of the dynamical system (\ref{sis1})-(\ref{sysLHS2}) for $a^2=0.5$ with the exact solution (orange line); {\bf b)} The phase portrait fragment on the interval $Z\in\left[-0.1, 1.5\right]$ and $X\in\left[-1.5, 1.5\right]$.}
\end{figure}

\begin{figure}[h!]
	\centering
	\includegraphics[width= 7.55cm]{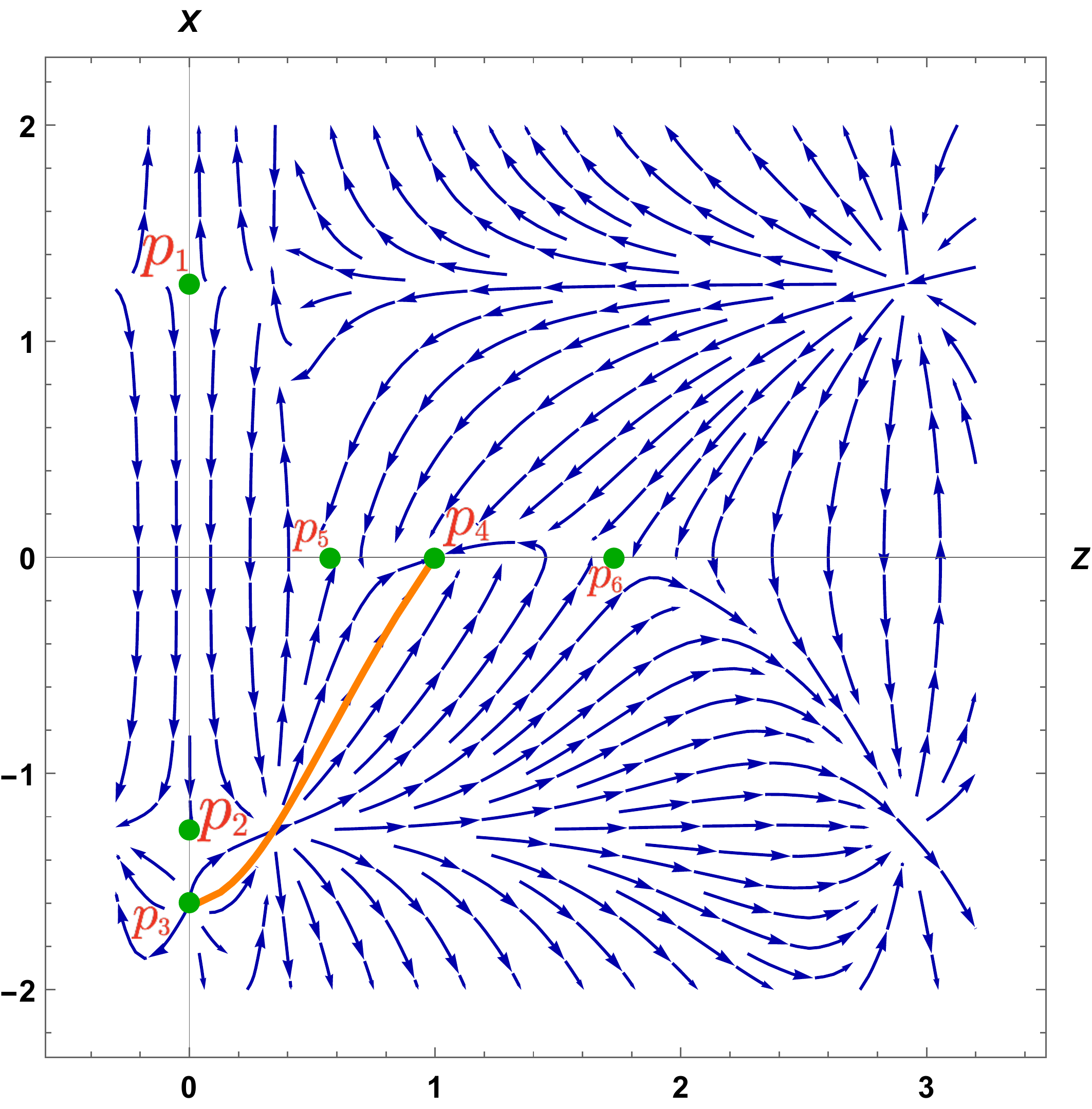}{\bf a)}
	\includegraphics[width= 7.55cm]{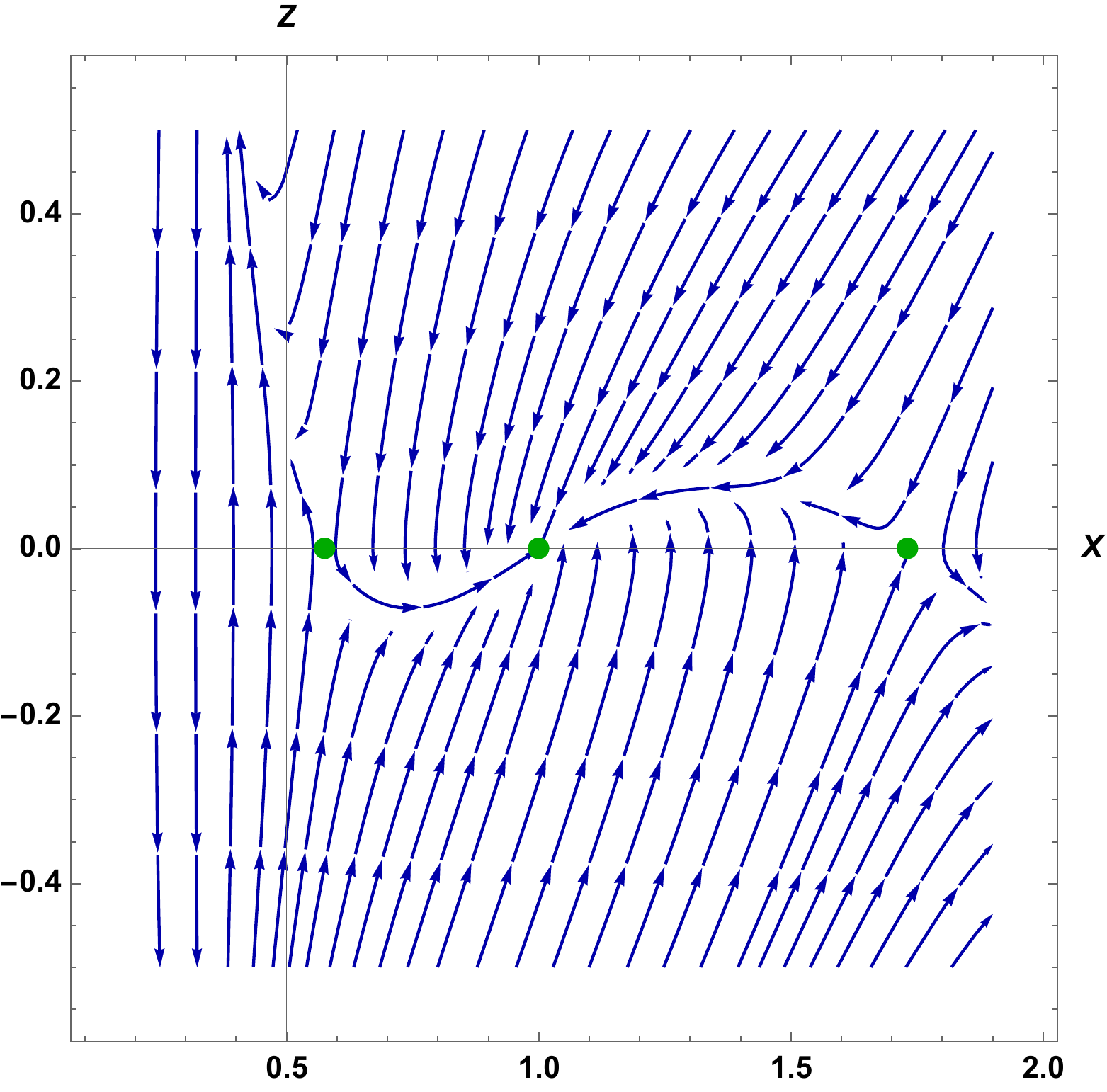}{\bf b)}
	\caption{\small {\bf a)} The phase portrait of the dynamical system (\ref{sis1})-(\ref{sysLHS2}) for $a^2=0.8$ with the exact solution (orange line); {\bf b)} The phase portrait fragment on the interval $Z\in\left[0.2, 1.9\right]$ and $X\in\left[-0.5, 0.5\right]$}
	\label{Fig:fig5}
\end{figure}

	\begin{figure}[h!]
	\centering
	\includegraphics[width= 7.55cm]{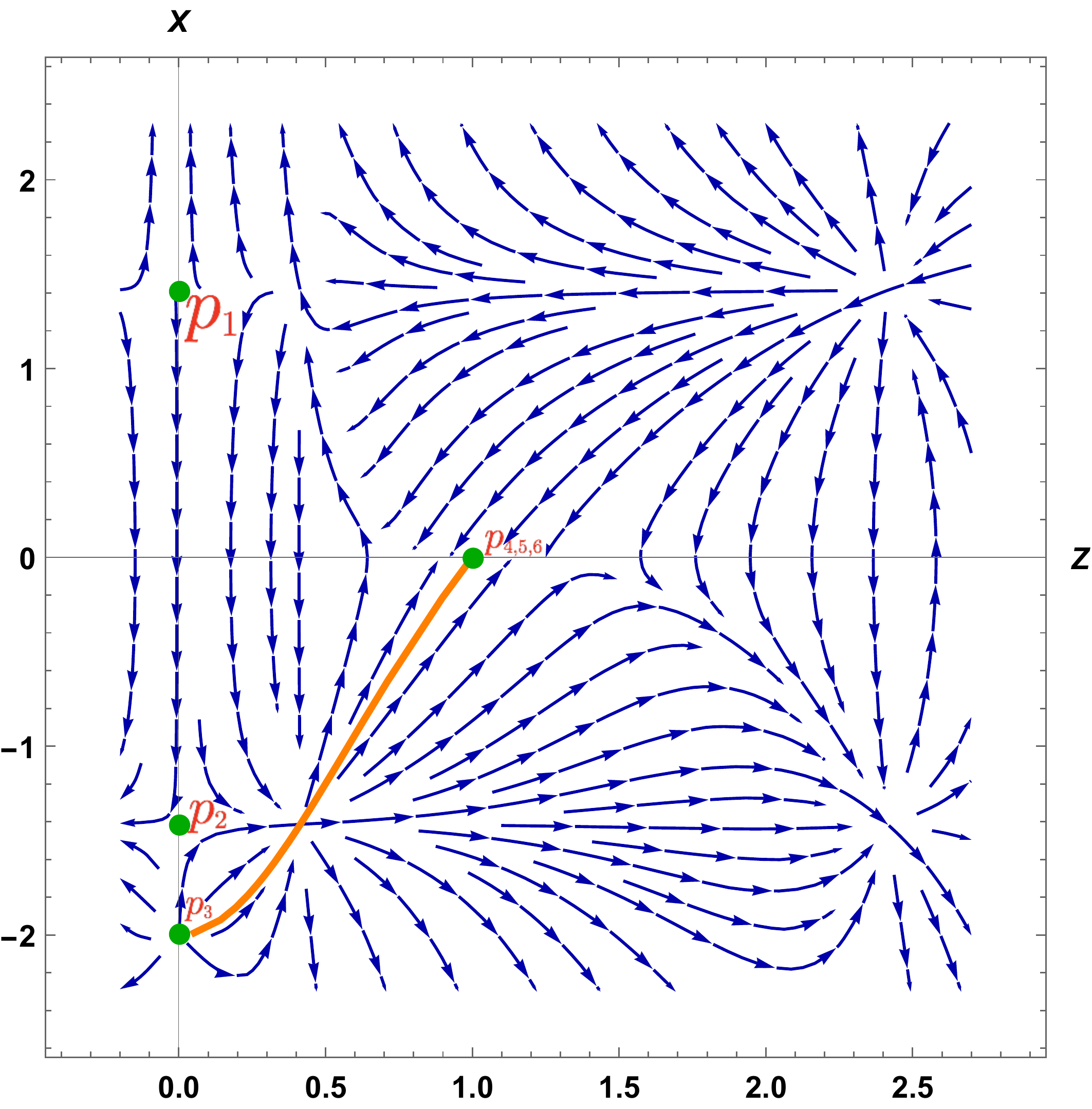}{\bf a)}
	\includegraphics[width= 7.55cm]{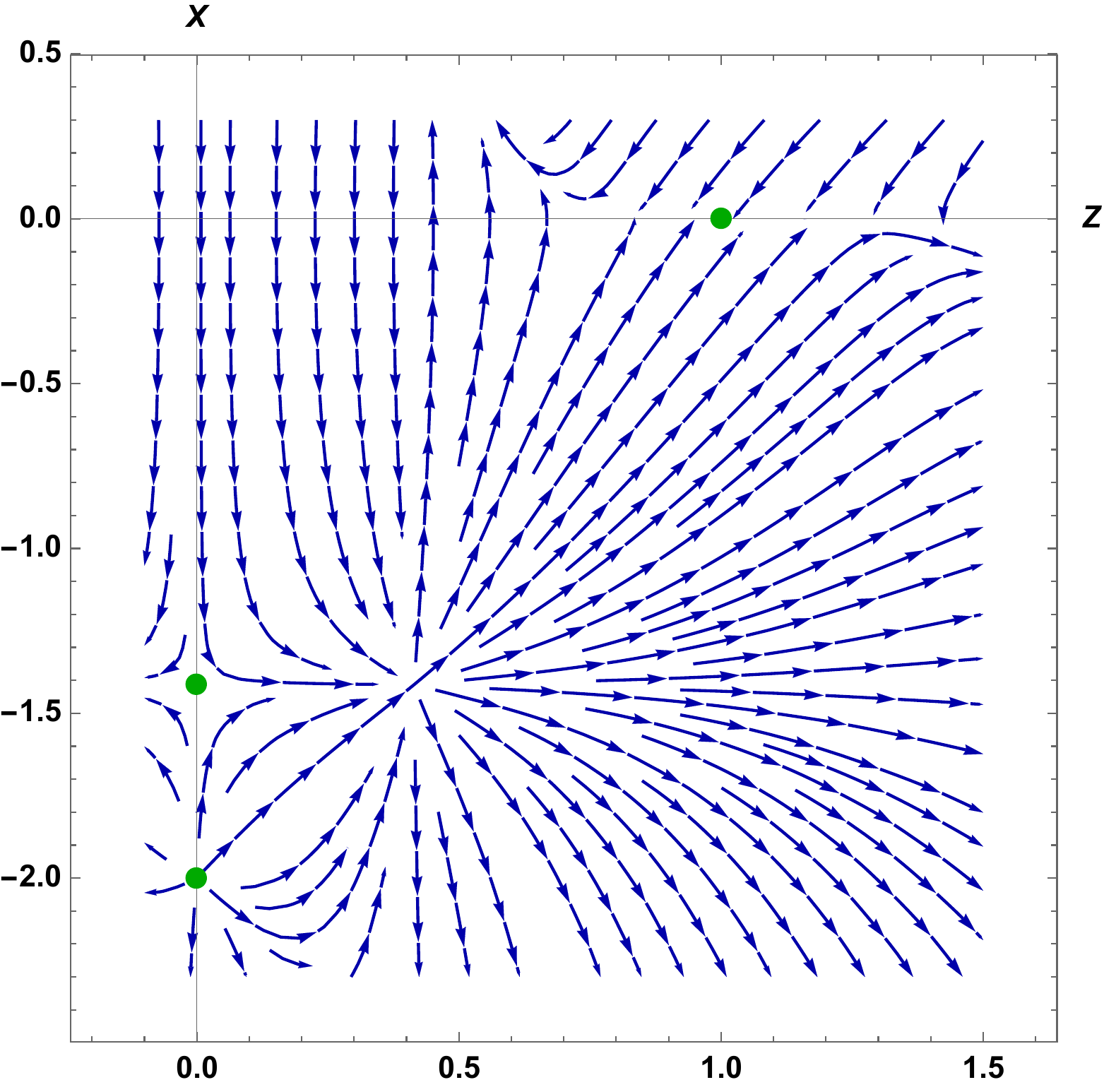}{\bf b)}
	\caption{\small {\bf a)} The phase portrait of the dynamical system (\ref{sis1})-(\ref{sysLHS2}) for $a^2=1$ with the exact solution (orange line); {\bf b)} The phase portrait fragment on the interval $Z\in\left[-0.1, 1.5\right]$ and $X\in\left[-2.3, 0,3\right]$.}
\label{fig:Fugure6}
\end{figure}

The case $a^2 =0.8$ ({\bf  Fig.5, a)}) is particularly interesting since all six points are represented on the phase portrait. Three points {\bf 1}, {\bf 2} and {\bf 3} have the same coordinate $Z_c=0$ and differ in value of $X_c$, while other three points {\bf 4}, {\bf 5} and {\bf 6} have different values of $Z_c$ and the same values of $X_c=0$.  In {\bf Fig.5 b)} we show a zoomed region of the points with $X_c=0$, one can see that there are phase trajectories between them.
 
 The phase portrait of the system for $a^2 =1$ is represented in {\bf Fig.6}. 
 We see 4 equilibrium points. The point  {\bf 4} with  $Z_c=1,X_c=0$, the points {\bf 1} and {\bf 2} with coordinates $Z_c =0$, $X_c=\pm\sqrt{2}$, and the point {\bf 3} with $Z_c =0$, $X=2$.
 
 The orange line in the {\bf Figs.} {\bf 4, 5, 6} depicts the flow, which was built by means of the exact solution (\ref{scafieldDeg})-(\ref{metricDeg}), found in \cite{Deger:2002hv}. The given flow always starts at the point $Z_c=1$,$X_c=0$ and  goes into the point with coordinate $Z_c=\infty$, which is, unfortunately, could not be represented on these phase portraits.

\setcounter{equation}{0}
\section{Asymptotic solutions to EOM and holographic RG flows}

The coordinates $Z_{c}$ and $X_{c}$  allow us to find asymptotic solutions near the critical points, i.e. one can write down the asymptotic behaviour of the scalar field and the metric. However, we also have to check if they obey  the original equations of motion (\ref{1T0})-(\ref{4T0}).

\subsection{General asymptotic solutions}

The obtained critical points may be divided into two types according to the value of the coordinate $Z$.  For $Z_c = 0$ the scalar field  goes to infinity, $\phi\to +\infty$, while  the scalar field has some constant value for non-zero  $Z_c$.

To find the form of the scale factor $A$ near a critical point with  $X_c$  we have to solve the equation (\ref{pr2}):
\be
\frac{d\dot{A}}{dw}\frac{1}{\dot{A}^2}=-\frac{X^2_c}{a^2},
\ee
which gives us
\be
\frac{1}{\dot{A}}  = \frac{\dot{A}_{0}X^{2}(w-w_0) +a^2}{\dot{A}_{0}a^2},
\ee
where $A_0$ and $w_0$ are constants of integration. Integrating the last equation, we obtain the general form of the scale factor:
\be\label{eqeq}
A = \frac{a^2}{X^2_{c}}\ln\Bigl[\frac{X^2_{c}\dot{A}_{0}(w-w_0)+ a^2}{X^2_{c}\dot{A}_{0}(w_1-w_0)+ a^2}\Bigr]+A_0.
\ee
 The dilaton can be restored using the  relation between $X_c$ and the  derivatives of the scalar field and the scale factor  (\ref{XvarDef}). So we get
\be\label{eded}
\phi = \frac{a^2}{X_{c}} \ln\Bigl[\frac{\dot{A}_{0}X^2_{c}(w-w_0)+a^2}{\dot{A}_{0}X^2_{c}(w_1-w_0)+a^2}\Bigr]+\phi_0,
\ee
where $\phi_0$ is a constant of integration.

Thus, taking into account (\ref{eqeq}) and (\ref{eded}), we find that the scale factor $A$ has the following asymptotic behavior near $Z_c=0$ for all $a$
\be
A\to -\infty,\quad \text{as}\quad w\to w_0-\frac{a^2}{X^2_c\dot{A_0}}.
\ee

Correspondingly, the metric  (\ref{metricT}) takes to the form
\be\label{genMetDW}
ds^2\cong\Bigl|\frac{\dot{A}_{0}X^2_{c}(w-w_0)+a^2}{\dot{A}_{0}X^2_{c}(w_1-w_0)+a^2}\Bigr|^{\frac{2a^2}{X^2_c}}(- dt^2+dx^2)+ dw^2.
\ee

\subsubsection{Asymptotic solutions with $Z_c =0$}
{\bf 1.} The equilibrium point $Z_c=0,\quad X_c=a\sqrt{2}$.\\

Plugging $Z_c=0,\quad X_c=a\sqrt{2}$ into (\ref{genMetDW}) we find the metric near this critical point
\be
ds^2\cong\bigg|\frac{2\dot{A}_0(w-w_0)+1}{2\dot{A}_0(w_1-w_0)+1}\bigg|(-dt^2+dx^2)+dw^2
\ee
and  the dilaton (\ref{eded}) 
\be\label{dilsecond}
\phi=\frac{a}{\sqrt{2}}\ln\bigg|\frac{2\dot{A}_0(w-w_0)+1}{2\dot{A}_0(w_1-w_0)+1}\bigg|+\phi_0.
\ee

Since $Z_c=0$, the scalar field goes to  $+\infty$,  so $a>0$ and $w\to w_0-\frac{1}{2\dot{A}_0}$,
or $a<0$ and $w\to +\infty$.

 It is useful to estimate the behavior of the potential for $\phi\to+\infty$
\be\label{Vasymp}
V \sim \frac{1}{2}\Lambda_{uv} e^{2\phi } (\frac{1}{4} (1-2 a^2)(4+e^{2\phi }) +2 a^2),
      \ee
thus  
 \be\label{VfirstExp}
 V \sim
 \begin{cases}
 -\infty, \quad\textrm{for}\quad 0\leq a^2\leq\frac{1}{2} ,\\
  +\infty, \quad\textrm{for}\quad a^2>\frac{1}{2}.
 \end{cases}
 \ee
Substituting  the scale factor and the dilaton (\ref{dilsecond}) into eqs.(\ref{1T0})-(\ref{4T0}) we get the following constraints:
\be\label{Vcondit2}
V=0,\quad
\frac{dV}{d\phi}=0,
\ee
which cannot be satisfied for any   $a$.

{\bf 2.} The equilibrium point $Z_c=0,\quad X_c= -a\sqrt{2}$.

The metric (\ref{genMetDW}) near this critical point in the domain wall coordinates can be written as follows
\be\label{metfirst}
ds^2\cong\bigg|\frac{2\dot{A}_0(w-w_0)+1}{2\dot{A}_0(w_1-w_0)+1}\bigg|(-dt^2+dx^2)+dw^2
\ee
and the corresponding scalar field (\ref{eded}) takes the form
\be\label{dilfirst}
\phi= - \frac{a}{\sqrt{2}}\ln\bigg|\frac{2\dot{A}_0(w-w_0)+1}{2\dot{A}_0(w_1-w_0)+1}\bigg|+\phi_0.
\ee
 
 As in the previous case of $Z_c=0$, i.e. $\phi\to +\infty$, that can be reached for $a<0$ with $w\to w_0-\frac{1}{2\dot{A}_0}$, and also for $a>0$ with $w\to +\infty$.

The asymptotic behavior of the potential coincides with the previous case (\ref{Vasymp}). So the solution (\ref{metfirst})-(\ref{dilfirst}) yields the constraint for the potential and its derivative (\ref{Vcondit2}), which
cannot be resolved for any $a$.

{\bf 3.} The equilibrium point $Z_c=0$, $X_c=-2a^{2}$.\\

Plugging the values of $Z_c$ and $X_c$ into  (\ref{genMetDW}) we find
\be\label{metricHSV}
ds^2\cong\bigg|\frac{4a^2\dot{A}_0(w-w_0)+1}{4a^2\dot{A}_0(w_1-w_0)+1}\bigg|^{\frac{1}{2a^{2}}}(-dt^2+dx^2)+dw^2
\ee
 and the dilaton has the form
\be\label{thirddil}
\phi=-\frac{1}{2}\ln{\Big|\frac{4a^2\dot{A}_{0}(w-w_0)+1}{4a^2\dot{A}_0(w_1-w_0)+1}\Big|}+\phi_0.
\ee
Since $Z_c=0$  the scalar field $\phi\to+\infty$, that is relevant for all  $a$ with $ w\to w_0-\frac{1}{4a^2 \dot{A}_0}$ \footnote{One can choose $w_{0} =w_{1}=\frac{1}{4a^2\dot{A}_{0}}$.}.
For $\phi \to +\infty$, the potential has the form (\ref{Vasymp})
and behaves as
\be
V\sim
\begin{cases}
 - \infty, \quad\textrm{for}\quad 0\leq a^{2}\leq\frac{1}{2},\\
 + \infty,\quad\textrm{for}\quad a^{2}>\frac{1}{2}.
\end{cases}
\ee
   
  Substituting the scale factor and the scalar field into eqs. (\ref{1T0})-(\ref{4T0})  we see that the solution exists for all $a$.\\
   Thus, for any $a$ the asymptotic solution (\ref{metricHSV})-(\ref{thirddil}) describes a supersymmetric non-conformal vacuum and corresponds to the solution (\ref{scafieldDeg})-(\ref{metricDeg}) near $w=0$.

\subsubsection{Asymptotic solutions with $Z_c=\textrm{const}$}

{\bf 4.} The equilibrium point $Z_c=1,\quad X_c=0$.

Since $Z_c=1$, then due to  (\ref{03}) the scalar field is 
\be\label{DilF}
\phi=0,
\ee
and the potential (\ref{DilPot}) has the following value at this point
 \be\label{Vsol}
 V= 2\Lambda_{uv}. \ee 
 One can find the asymptotic solution for the scale factor from the Einstein equation (\ref{1T0})
 \be
 2\dot{A}^2= -V,
 \ee
taking into account (\ref{Vsol}),  we find 
\be
A= \sqrt{-\Lambda_{uv}}\left(w-w_0\right),
\ee
where $w_{0}$ is a constant of integration.

Then the metric (\ref{metricT}) near the point $Z_c=1, X_c=0$ has the form
\be\label{AdSF}
ds^2 \approx e^{2\sqrt{-\Lambda_{uv}}\left(w-w_0\right)}\left(-dt^2+dx^2\right)+dw^2.
\ee
Note that the metric (\ref{AdSF}) is nothing else but the $AdS_{3}$ metric in the domain wall coordinates.
The scalar field (\ref{DilF}) and the metric (\ref{AdSF})  solve eqs. (\ref{1T0})-(\ref{4T0})  for all $a$.
The asymptotic metric (\ref{AdSF})  with $\phi =0$ represents a supersymmetric AdS vacua for any $a$. Moreover,  it corresponds to the solution (\ref{scafieldDeg})-(\ref{metricDeg}) with $w \to \infty$.

{\bf 5.} The equilibrium point $Z_c=\sqrt{\frac{1-2|a| \sqrt{1-a^2}}{2 a^2-1}},\quad X_c=0$.

From (\ref{03})
we find that the dilaton near this point is given by
\be\label{DilT}
\phi = \ln\left(\sqrt{\frac{1-2|a| \sqrt{1-a^2}}{2 a^2-1}}\right),
\ee
that corresponds to  the potential
\be
V=\frac{2 a^4 \Lambda_{uv} }{2 a^2-1}.
\ee
The scale factor can be find from (\ref{1T0})
\be\label{point5SCF}
A= a^{2} \sqrt{-\frac{\Lambda_{uv}}{2a^2- 1}}(w-w_0),
\ee
where $w_{0}$ is a constant of integration.

Then the metric (\ref{metricT}) near the point $Z_c=\sqrt{\frac{1-2|a| \sqrt{1-a^2}}{2 a^2-1}}, X_c=0$ is given by
\be\label{AdST}
ds^2 \approx e^{2a^{2} \sqrt{-\frac{\Lambda_{uv}}{2a^2-1}}(w-w_0)}\left(-dt^2+dx^2\right)+dw^2,
\ee
which is again $AdS_{3}$.
Note that the solution (\ref{DilT}) and (\ref{AdST}) solves EOM (\ref{1T0})-(\ref{4T0})  for $a^2>\frac{1}{2}$.

The solution (\ref{DilT}), (\ref{point5SCF}) with $a^2>\frac{1}{2}$ describes a non-supersymmetric AdS vacua and corresponds to the extremum of the potential (\ref{ScFConst}) with $\phi_{3} =  \frac{1}{2}\ln\left(\frac{1- 2|a|\sqrt{1-a^{2}}}{2a^2-1}\right)$.

{\bf 6.} The equilibrium point $Z_c=\sqrt{\frac{1+2|a|\sqrt{1-a^2}}{2a^2-1}},\quad X_c=0$.

As in the previous case, the dilaton near this point takes a constant value:
\be\label{scalarF6}
\phi = \ln\left(\sqrt{\frac{1+ 2 |a|\sqrt{1-a^2}}{2 a^2-1}}\right).
\ee
 Similarly, the potential of the scalar field also has a constant value and it is defined by the parameter $a$:
\be
V=\frac{2 a^4 \Lambda_{uv} }{2 a^2-1}.
\ee
The scale factor is
\be\label{point6SCF}
A= a^{2} \sqrt{-\frac{\Lambda_{uv}}{2a^2-1}}(w-w_0),
\ee
where $w_{0}$ is a constant of integration.

Therefore the metric (\ref{metricT}) near the point $Z_c=\sqrt{\frac{1+ 2 \sqrt{a^2-a^4}}{2 a^2-1}}, X_c=0$ has the following form
\be\label{metric6}
ds^2 \approx e^{2a^{2} \sqrt{-\frac{\Lambda_{uv}}{2a^2-1}}(w-w_0)}\left(-dt^2+dx^2\right)+dw^2,
\ee
that corresponds to the $AdS_3$ spacetime in the domain wall coordinates.

The metric (\ref{metric6}) and the scalar field (\ref{scalarF6}) are solutions to eqs.(\ref{1T0})-(\ref{4T0}) for  $a^2>\frac{1}{2}$.
The solution (\ref{scalarF6}), (\ref{point6SCF}) with $a^2>\frac{1}{2}$ is a non-supersymmetric AdS vacua and corresponds to the extremum of the potential (\ref{ScFConst}) with $\phi_{2} =  \frac{1}{2}\ln\left(\frac{1+ 2|a|\sqrt{1-a^{2}}}{2a^2-1}\right)$.

\subsection{ Holographic RG flows}

Here we briefly summarize our results for the asymptotic solutions of our model:
\begin{itemize}
    \item the supersymmetric solution with an $AdS_3$ asymtotics and $\phi =0$ which is valid for all $a$ and corresponds to the point {\bf 4} $Z_c=1$, $X_c=0$. 
    \item  the non-supersymmetric solutions with the $AdS_3$ asymtotics and constant scalar fields $\phi= \ln\left(\sqrt{\frac{1\pm 2 |a|\sqrt{1-a^2}}{2 a^2-1}}\right)$, which are valid for $a^2>\frac{1}{2}$, and correspond to the equilibrium points {\bf 5,6} with $Z_c=\sqrt{\frac{1\pm 2 |a|\sqrt{1-a^2}}{2 a^2-1}}$, $X_c=0$. 
    \item  there are also asymptotic supersymmetric solutions with hyperscaling violating metrics and the dilaton tending to $\pm \infty$ (depending on $a$), which correspond to the point {\bf 3} with  $Z_c=0$, $X_c=-2a^2$; the solutions are valid for any $a$;
    \item  the asymptotic solutions for the scalar field and the metric, corresponding to the equilibrium points  {\bf 1} and {\bf 2}, do not fulfill the equations of motion for all values of $a$.
  \end{itemize}
  
  We also note that we reproduced the exact solution \cite{Deger:2002hv} on the phase portraits ~\ref{fig:Fugure3}--\ref{fig:Fugure6}. One can see that it starts from the critical point {\bf 4}, $AdS_3$ with $\phi=0$, and flows to $\phi\to+\infty$.

  According to the holographic dictionary,   the energy scale of the dual field theory is associated to the scale factor of the gravity solution, i.e.
\be
\mathcal{A}=e^{A}.
\ee
At the same time, we assume that the  RG flow starts at fixed point with high energy (UV) and flows to another fixed point with small energy (IR). Thus, the stability type of the critical points must be consistent with the energy scale.

In table ~\ref{tab:2} we present the characteristics of the critical points of the dynamical system, which obey the equations of motion (\ref{1T0})-(\ref{4T0}). We also take into account the direction of the energy scale and give an interpretation in terms of $UV/IR$ fixed points.

\begin{table}[t]
{\noindent{ \extrarowheight = 2pt 
\centering
\begin{tabular}{|c|c|c|c|}
\hline     &  $V(\phi)$                                                                       & \begin{tabular}[c]{@{}c@{}}{\bf Type according to} \\ {\bf energy scale}\end{tabular}&UV/IR \\ \hline
\multirow{2}{*}{$p_3$} &  $V\to-\infty$, $a^2\in (0;\frac{1}{2}) $  &  unstable (saddle,$\,a^2\in (0;\frac{1}{2})$ ) &\multirow{2}{*}{IR}\\
  & $V\to+\infty$, $\,a^2\in (\frac{1}{2};1] $  & stable (stable node,$\, a^2\in (\frac{1}{2};1]$ )  &\\ \hline
\multirow{2}{*}{$p_4$}  & \multirow{2}{*}{${\textrm{const}}$}      & \multirow{2}{*}{unstable (unstable node for all  $a$)}  & \multirow{2}{*}{UV} \\
&      &   &   \\ \hline
\multirow{2}{*}{$p_5$}    & \multirow{2}{*}{${\textrm{const}}$}                                                       & \multirow{2}{*}{unstable (saddle for all $a$)}  & \multirow{2}{*}{IR/UV, $a^2\in(\frac{1}{2};1)$}  \\
   &                  &   &   \\ \hline
\multirow{2}{*}{$p_6$} &  \multirow{2}{*}{${\textrm{const}}$}                                                              & \multirow{2}{*}{unstable (saddle for all $a$)} &\multirow{2}{*}{ IR/UV, $a^2\in(\frac{1}{2};1)$} \\
& &           &     \\ \hline
\end{tabular}

}}
\caption{\small The fixed points and their characteristics.}
\label{tab:2}
\end{table}

The possible holographic RG flows can be represented as follows:
\begin{itemize}
    \item for $a^2\leq \frac{1}{2}$:
    \begin{enumerate}
    \item the flow starts at the unstable supersymmetric UV fixed point ($p_4$), which corresponds to an $AdS$ boundary and  $\phi=0$ and goes to the unstable supersymmetric IR fixed point ($p_3$) related to the hyperscaling violating metric and the dilaton $\phi\to +\infty$, $V \to -\infty$; this flow corresponds to the exact solution (\ref{scafieldDeg})-(\ref{scafDeg}). Assuming the Dirichlet boundary conditions, this flow with $a^2\leq\frac{1}{2}$ can be associated with a deformation of the CFT by a relevant operator, while the flow with  $a^2=\frac{1}{2}$ is driven by a non-zero vev of the scalar operator.
        \end{enumerate}
    \item $a^2>\frac{1}{2}$: \begin{enumerate} 
    \item the RG flow starts at the unstable asymptotically AdS UV fixed point ($p_4$) with $\phi=0$ and flows to the stable IR fixed point ($p_3$) with $\phi\to +\infty$ and $V(\phi)\to +\infty$; this flow corresponds to the exact one from \cite{Deger:2002hv} and seems to be singular since the potential is unbounded from above. Therefore, the solution with the Dirichlet boundary conditions can describe  a deformation by a nonzero VEV of the operator $\mathcal{O}_{\phi}$.
    \item  the flow begins at the unstable supersymmetric UV fixed point ($p_4$), with the $AdS_3$ boundary and  $\phi=0$ and flows to the unstable non-supersymmteric IR fixed point $p_5$ ($p_6$) with an asymptotically AdS metric and the constant scalar field. Note that this flow was discussed in \cite{ParkRoLee}. The solutions with the Dirichlet conditions can be associated to a deformation triggered by a relevant operator.
   \item the flow can start from the unstable non-supersymmetric AdS fixed point $p_5$ ($p_6$) and goes to the stable IR ($p_3$, $\phi\to +\infty$).
    \end{enumerate}
\end{itemize}

\subsection{Bifurcations in the model}

It is interesting to study bifurcations in our model. Since the system contains an external parameter ($a$), one can expect that a small change of it may give rise to a change of the phase portrait.

Note that bifurcations are characterized by a vanishing eigenvalue of the Jacobian matrix. Moreover, some types of bifurcations require that the determinant of the Jacobi matrix vanishes at the bifurcation point \cite{Gukov}. 
 
\begin{figure}[h!]
	\centering
	\includegraphics[width= 7.5 cm]{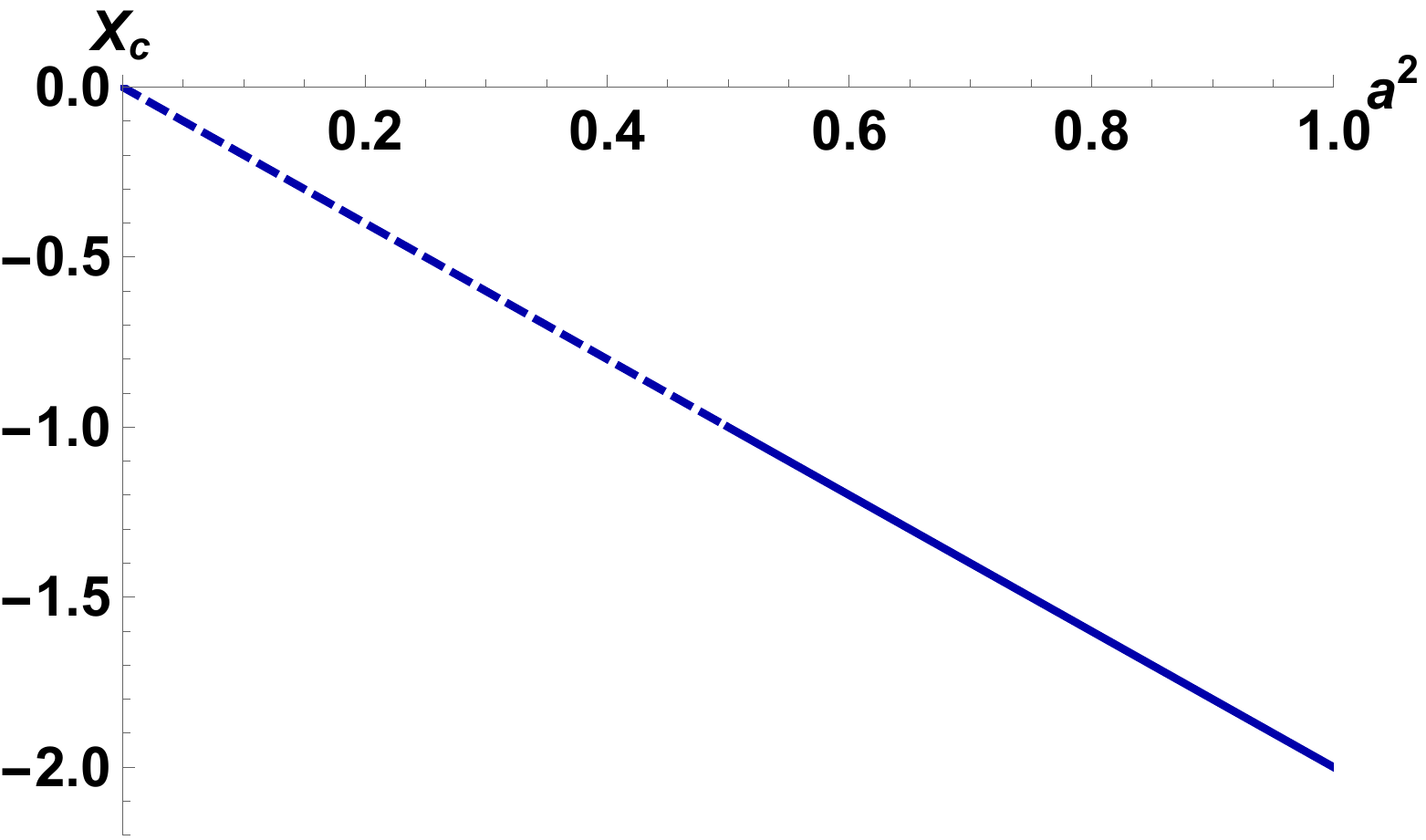}
	\caption{The bifurcation diagram for  $X_c=-2a^2$, the unstable points are on the dashed line, while the stable are on the solid one.}
	\label{Fig:BifrD}
	\end{figure}

The dynamical system  (\ref{sis1})-(\ref{sysLHS2}) with $Z_c =0$ takes the form:
\be\label{BIFYZ0}
\dot{X} = \left(\frac{X^2}{a^2} -2\right)\left(X + 2a^2\right).
\ee
The equilibria of (\ref{BIFYZ0}) are 
\be\label{eqlpBif}
{\bf 1)} X_c = \sqrt{2}a,\quad {\bf 2)}X_c = -\sqrt{2}a,\quad {\bf 3)} X_c = -2a^2.
\ee
For the cases {\bf 1)} and {\bf 2)} the qualitative behavior of the solutions changes at $a=0,\pm\frac{1}{\sqrt{2}}$,  see table~\ref{TableTwo}. The asymptotic solutions near these critical points don't satisfy eqs. (\ref{1T0})-(\ref{4T0}) and we omit their consideration.  However, the last case  with $X_{c}=-2a^2$ (\ref{eqlpBif}) is also interesting. So, taking into account the direction of the energy scale, the points $X_c=- 2a^2$ are unstable (saddles) for $a^2\in (0;\frac{1}{2})$, while they are stable for $a^2\in(\frac{1}{2};1)$ (stable nodes). So, the values $a=\pm\frac{1}{\sqrt{2}}$
are the bifurcation points. At these values of $a$ one of eigenvalues (\ref{eigenC3}) vanishes and the determinant is also equal to zero. We plot the bifurcation diagram in Fig.~\ref{Fig:BifrD}. Note, that for $a=0$ we have one zero eigenvalue and the vanishing determinant of the Jacobian matrix as well. Calculating coordinates of all points (\ref{eqlpBif}), one finds that they collide at $a=0$.\\

For $X_c=0$ we also have three critical points:
\be
Z_c =0,\quad Z_{c}=\sqrt{\frac{1+2|a|\sqrt{1-a^2}}{2a^2-1}},\quad Z_{c}=\sqrt{\frac{1-2|a|\sqrt{1-a^2}}{2a^2-1}}.
\ee
These points collide at $a^2=1$, so one of their eigenvalues and the determinant of the Jacobi matrix tend to be zero.

It is said, an operator, which has a vanishing eigenvalue, crosses the marginal bound. 
Indeed,  for $a^2=1$ the scaling dimensions (\ref{DeltaOp}),(\ref{DeltaOp2}) of the operators show that they are marginal and for $1<a^2 <\frac{1+\sqrt{2}}{2}$ are irrelevant. 

In Sect.~\ref{Sect:3} we have calculated the conformal dimension of the operator $\Delta = 1+|1- 2a^2|$ for different values $a$. It is suggested that the behavior of the determinant of the Jacobian matrix is equivalent to $\Delta-d$   \cite{Gukov}, that for our case:
\be\label{marigcross}
\Delta  - d  = |1- 2a^2| -1.
\ee 

For different values of $a^2$ we have the following behaviours for the quantity (\ref{marigcross}):

\begin{enumerate}
\item  for $a^2=0$ and $a^2=1$ we have $\Delta-d=0$;
\item for $a\in (-\frac{1}{\sqrt{2}};0)\cup(0;\frac{1}{\sqrt{2}})$, we get $\Delta -d = -2a^2$;
 \item for $a^2=\frac{1}{2}$, we have $\Delta-d=-1$;
 \item for $a\in(-1;-\frac{1}{\sqrt{2}})\cup (\frac{1}{\sqrt{2}};1)$ we get  $\Delta-d=2(a^2-1)$.
\end{enumerate}

 \section{Conclusions}
 
 In this paper, we have studied holographic RG flows in a 3d supergravity model with the scalar field   and the potential
 \cite{Sezgin,Deger:2002hv} from the side of the dynamical system theory. 
 
 The equations of motion have been reduced to the autonomous dynamical system by introducing new variables $Z$ and $X$. Note that these variables  express  behaviors of the potential and the holographic  $\beta$-function. It is supposed that the stationary points of the dynamical system, which obey the original equations of motion, can be related to fixed points of a dual field theory.
  For our autonomous system we have found  stationary points (3 points), which are either characterized by a constant value of the scalar field or the scalar field tending to infinity (3 points). Linearizing the dynamical system, the equilibrium points have been analyzed for stability. We have found that a type of stability of stationary points with $\phi\to +\infty$ may depend on the parameter $a$, which determines the behavior of the dilaton potential (see table~\ref{TableTwo}). For $a^2=0,\frac{1}{2},1$ we are not able to find out a type stability since one of the eigenvalues becomes equal to zero.  
  The phase portraits for the system (\ref{sis1})-(\ref{sysLHS2}) have been presented in Figs. ~\ref{fig:Fugure3}--\ref{fig:Fugure6} for different values of $a$. in Figs. ~\ref{fig:Fugure3}--\ref{fig:Fugure6} {\bf a)} we have also shown phase trajectories corresponding to the exact solution found in \cite{Deger:2002hv}.
  
  Using coordinates of the critical points ($Z_c$,$X_c$) we have reconstructed asymptotic solutions for the metric and scalar field near these points.
  As expected, the metrics near the critical points, corresponding to a constant scalar field, are asymptotically AdS spacetimes, while near the critical points with the scalar field tending to infinity we have obtained metrics with hyperscaling violation. We have also found that solutions near two critical points (with $\phi\to +\infty$) don't fulfill the equations of motion. Moreover, for two of three asymptotically AdS solutions we have found the constraint on the parameter: $a^2>\frac{1}{2}$, so these solutions correspond to local minima of the potential, see Fig.~\ref{Fig:DilPot} {\bf a)}.
  
  We have presented possible holographic RG flows taking into account the direction of the energy scale. It's interesting to note that all UV fixed points, which correspond to asymptotically AdS solutions, are repulsive. For $a^2\leq\frac{1}{2}$ there is a supersymmetric flow between an $AdS_{3}$ (UV) to  a  hyperscaling violating metric (IR) with $\phi\to+\infty$ ($V\to -\infty$) (\ref{metricHSV}), which for $a^2 < \frac{1}{2}$ corresponds to a deformation of a dual CFT by a relevant operator, while for $a^2=\frac{1}{2}$ is triggered by a non-zero vev of the scalar operator.  As for $a^2>\frac{1}{2}$ the potential has additional extrema, which yield to  non-supersymmetric AdS solutions.
  So one can observe a holographic RG flow between two AdS fixed points, see Fig.~\ref{Fig:fig5}{\bf b)}, which can be associated to a deformation by a relevant operator. This case was explored in \cite{ParkRoLee}. There can also be  flows from non-supersymmetric $AdS_{3}$ (UV) fixed points to an IR fixed point related to a hyperscaling violating metric. However,  in this case the potential is unbounded from above \cite{Gubser:2000nd}. 
    
  For IR fixed points, which are described by metrics with hyperscaling violation, we have found an interesting observation that  stability of the points changes if the parameter $a$ belongs to different regions, i.e. there is a bifurcation point at $a^{2}=\frac{1}{2}$. At this value of $a$ the potential changes its asymptotic behavior, see Fig.~\ref{Fig:DilPot}.
  
  It wound be very interesting to perform a careful analysis of  the holographic RG flows, which we have constructed, with Neumann and the Mixed boundary conditions \cite{Papadimitriou:2007sj}.
 
Another interesting direction for future research would be to explore  a finite temperature generalization of the model, i.e. to find finite temperature solutions and study its thermodynamic properties. In this case we expect to observe bifurcations related with phase transitions. A second direction concerns to a construction of an uplift to a higher-dimensional model to clarify dual CFT.

\paragraph{Data Availability Statement}\
	\\
	Data sharing not applicable to this article as no datasets were generated or analysed during the current study.

\section*{Acknowledgments}
We are grateful to I. Ya. Aref'eva, I. Bakhmatov, K. Gubarev, H. Dimov, E.Musaev for useful stimulating discussions and comments. The work is supported by Russian Science Foundation grant 20-12-00200. We also thank to the EPJ Plus
referee for careful reading of our paper and valuable comments.

\newpage
\appendix

\setcounter{equation}{0} \renewcommand{\theequation}{A.\arabic{equation}}

\section{Geometric characteristics of the metric and stress-energy tensor}

 The metric in the domain wall coordinates is described by the following expression:
\be\label{metricTA}
ds^2=e^{2A(w)}\left(-dt^2+dx^2\right)+dw^2.
\ee
Then non-zero Ricci tensor components take the form:
\be
R_{tt}= e^{2 A} ( \ddot{A}+2\dot{A}^2),\quad
R_{xx}=-e^{2 A}  (\ddot{A}+2 \dot{A}^2),\quad
R_{ww}=-2  (\ddot{A}+\dot{A}^2),
\ee
and the Ricci scalar is given by:
\be
R=-2 (2 \ddot{A}+3 \dot{A}^2).
\ee
The components of the Einstein tensor are:
\be
G_{tt}=-e^{2A}(\ddot{A}+\dot{A}^2),\quad G_{xx}=e^{2A}(\ddot{A}+\dot{A}^2),\quad G_{ww}=\dot{A}^2.
\ee
Stress-energy-momentum tensor, defined as 
\be
T_{\mu\nu}=\frac{1}{a^2}\left(\partial_{\mu}\phi\partial_{\nu}\phi-\frac{1}{2}g_{\mu\nu}\partial_{\sigma}\phi\partial^{\sigma}\phi\right)-\frac{1}{2}g_{\mu\nu}V,
\ee
will have the following non-zero components:
\bea
T_{tt}&=&\frac{1}{a^2}\left(-g_{tt}\dot{\phi}^2g^{ww}\right)-\frac{1}{2}g_{tt}V=\frac{e^{2 A}}{2}\left(\frac{\dot{\phi}^2}{a^2}+V\right),\\
T_{xx}&=&\frac{1}{a^2}\left(-\frac{1}{2}g_{xx}\dot{\phi}^2g^{ww}\right)-\frac{1}{2}g_{xx}V=-\frac{e^{2 A}}{2}\left(\frac{\dot{\phi}^2}{a^2}+V\right),\\
T_{ww}&=&\frac{1}{a^2}\left(\dot{\phi}^2-\frac{1}{2}\dot{\phi}^2\right)-\frac{1}{2}g_{ww}V=\frac{1}{2}\left(\frac{\dot{\phi}^2}{a^2}-V\right).
\eea

\end{document}